# Machining of Spherical Component Fabricated by Selected Laser Melting, Part II: Application of Ti in Biomedical

Amir Mahyar Khorasani

School of Engineering, Faculty of Science Engineering and Built Environment, Deakin University, Waurn Ponds, Victoria, Australia

a.khorasani@deakin.edu.au

## Abstract

Ti and Ti-Based alloys have unique properties such as high strength, low density and excellent corrosion resistance. These properties are essential for the manufacture of lightweight and high strength components for biomedical applications. In this paper, Ti properties such as metallurgy, mechanical properties, surface modification, corrosion resistance, biocompatibility and osseointegration in biomedical applications have been discussed. This paper also analyses the advantages and disadvantages of various Ti manufacturing processes for biomedical applications such as casting, powder metallurgy, cold and hot working, machining, laser engineering net shaping, superplastic forming, forging and ring rolling. The contributions of this research are twofold, firstly scrutinizing the behaviour of Ti and Ti-Based alloys in-vivo and in-vitro experiments in biomedical applications to determine the factors leading to failure, and secondly strategies to achieve desired properties essential to improving the quality of patient outcomes after receiving surgical implants. Future research will be directed toward manufacturing of Ti for medical applications by improving the production process, for example using optimal design approaches in additive manufacturing and investigating alloys containing other materials in order to obtain better medical and mechanical characteristics.

## Keywords

Biocompatibility, Bioimplants fabrication, Mechanical properties, Osseointegration, Ti-Based biomaterial

## Contents

**Glossary of Abbreviations and symbols**

Chapter 2

| | |
|---|---|
| CP | Commercially pure |
| CBN | Cubic boron nitride |
| MQL | Minimum quantity lubricant |
| HSM | High speed machining |
| CNC | Computer numerical control |
| HBN | Hexagonal boron nitride |
| HIPed | Hot isostatic pressed |
| NWG | Number of wheel grits |
| CAD | Computer aided design |
| CAM | Computer aided manufacturing |
| CVD | Chemical vapour deposition |
| PEEKs | Polyether ether ketone |
| $F_c$ | Cutting force |
| $F_t$ | Thrust force |
| $F_z$ | Feed force |
| DLC | Diamond-like carbon |
| PCBN | Polycrystalline cubic boron nitride |
| HA | Hydroxyapatite |

## 1.1  Introduction

After the first summit held on biomaterials at Clemson University, in the USA in 1969 biomaterials was introduced to the scientific society and received significant attention due to the potential for increasing people's health.  For example, total hip replacement is recommended for people who have medical issues related to excessive wear of the acetabular, osteoarthritis, accident or age. Researches have shown that about 230000 total hip arthroplasty (orthopedic surgery where the articular surface of a musculoskeletal joint is replaced, remodelled, or realigned) have been carried out annually in the USA and will increase in next few decades (1, 2). Due to some phenomena such as an absence of biological self-healing process, wear or excessive loading, degeneration occurs in human joints. It was reported that the number of people who suffer from these problems from 2002 to 2010 increased seven times. Based on Kutz's et al. (3) research in 2007 it was estimated that the demand of the hip and knee that are made by Co and Ti-Based alloys will increase 174% (57200 operations) and 673% (3.48 billon operations) respectively by the end of 2030 (4-8).

Artificial materials contain metals, ceramics, composites, polymers or natural materials which are used in the making of implants, structures or joints to replace the missing or diseased biological parts are called biomaterials. The use of biomaterials in these engineering and medical applications results in improved quality of human life as well as increased longevity. Biomaterials are used as various parts of the human body such as a stents in blood vessels, artificial knees, hips, elbows, dental applications, shoulders and valves in the heart. Figure 1 shows the demands of market for prosthetic joints made by Ti and Co-Based alloys in 2012 for Australia (9-15). Due to the deterioration of body parts by increasing human age the demand of using bioimplants has increased dramatically (16).

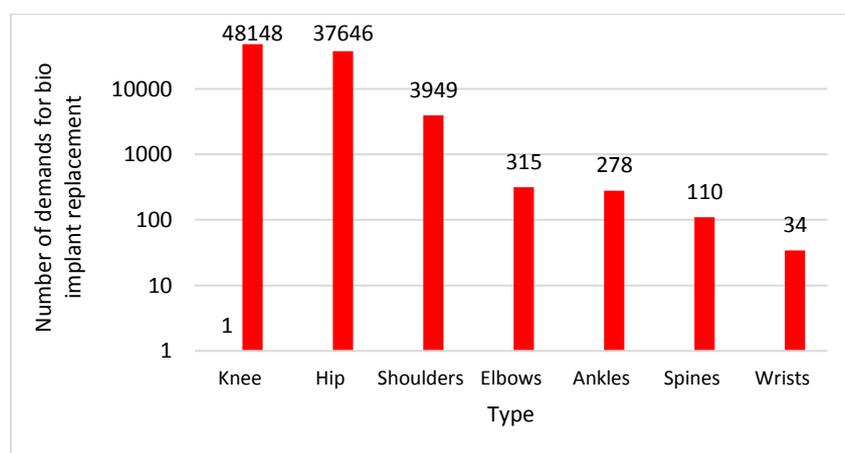

Figure 1 Number of demand for prosthetic bioimplants in Australia in 2012

Metal biomaterials are extensively used in medical applications due to their high strength and corrosion resistance; in addition stainless steel, Ti, magnesium and Co-Based alloys have superior biomedical properties among other metal biomaterials. Ti and Ti-Based alloys have excellent characteristics such as biocompatibility, osseointegration, high wear and corrosion resistance, low compatibility issues and high strength, thus recent attentions have been directed towards the development of these materials. Table 1 illustrates the properties of Ti among other light metals that are used in biomedical applications, it can be seen Ti has a high elastic modulus, melting and boiling point. The demand of using biomaterials is associated with different parameters, such as elasticity modulus, so alloys with higher strength have more usage in biomedical applications. Figure 2 shows the most common Ti and Ti-Based alloys that are used in biomedical application and their associated elasticity modulus (16-21).

Table 1 Physical properties of light metals used as biomaterials (22)

| Element Properties | Aluminium | Magnesium | Titanium |
|---|---|---|---|
| Melting Point $C^0$ | 660 | 650 | 1678 |
| Boiling Point $C^0$ | 2520 | 1090 | 3289 |
| Density g $cm^{-3}$ | 2.700 | 1.740 | 4.512 |
| Elastic Modulus GPa | 70 | 45 | 120 |
| Thermal Conductivity $Wm^{-1}k^{-1}$ | 238 | 156 | 26 |
| Hardness HBW | 160 | 44 | 716 |

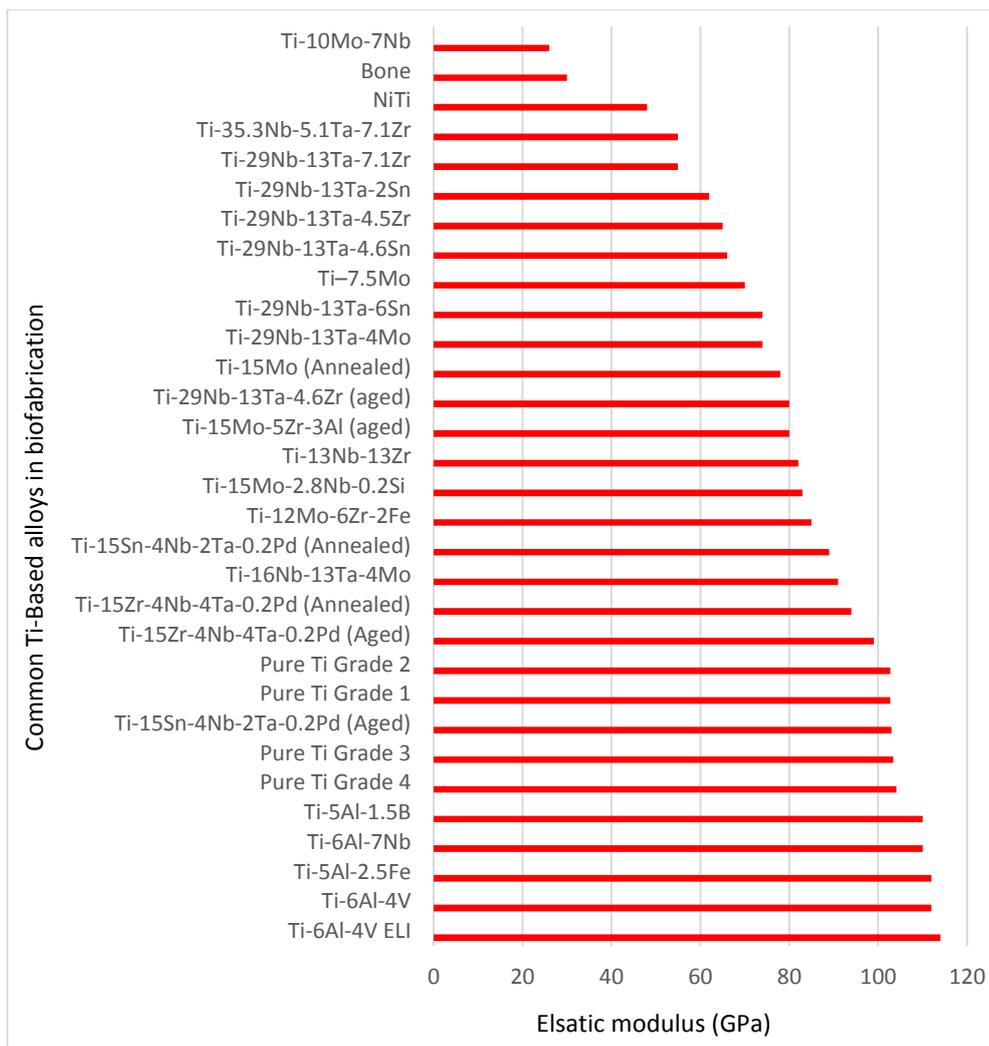

Figure 2 Common Ti-Based biomaterial elasticity modulus (9, 16-22)

Ti alloys especially in the α+β phase are used extensively in the human body due to their non-toxic and low allergenic properties, these give rise to a higher level of biocompatibility. Super elasticity and shape memory are also increasingly important characteristics not only in bio-applications, but also in different industries such as automotive and aerospace. Super elastic properties and shape memory of Ti are complicated, and therefore is fertile ground for research (23-31). Elements such as Nb, Mo, Sn, Ta and Zr are selected as the safest alloying metals in order to adjust properties of the biomaterial and maintain its suitability for implantation. There is 26 groups of Ti alloys that are used in biomedical application which are listed below: Ti, Ti-Al-B, Ti-Al-Nb, Ti-Al-Nb-Ta, Ti-Al-V, Ti–Fe–Ta, Ti–Mo, Ti–Mo–Al, Ti–Mo–Ga, Ti–Mo–Ge, Ti-Mo-Nb, Ti–Mo–Nb–Si, Ti–Mo–Zr–Al, Ti–Mo–Zr–Fe, Ti–Mo–Zr–Sn, Ti–Nb–Hf, Ti–Nb–Sn, Ti–Nb–Ta–Mo, Ti–Nb–Ta–Sn, Ti–Nb–Ta–Zr, Ti–Nb–Zr, Ti–Sn–Nb–Ta, Ti–Ta, Ti–Ta–Zr, Ti–Zr, Ni-Ti (10, 16, 32-35). Among these groups only Ti–Al–4V ELI and Ti–Al–7Nb have been standardized for biomaterials in ASTM (36).

As almost all of Ti-Based biomaterials that are used today for biomedical applications follow the (ISO 5832) standards that are listed as: (ISO 5832-2) unalloyed Ti, (ISO 5832-3) wrought Ti–6Al–4V alloy, (ISO 5832-11) wrought Ti-6Al-7Nb alloy, (ISO 5832-14) wrought Ti–15Mo–5Zr–3Al alloy. Furthermore, Ti has made a great contribution in dental applications such as removable prostheses, maxilla facial prostheses and implant supporting material due to superior biocompatibility, light weight, high wear resistance and the ability to manufacture implants with a high accuracy in order to fit with teeth and jaw bones (37-40).

The paper is divided into two main parts. The initial deals with properties such as metallurgical and mechanical characteristics together with a discussion fabrication processes and reviewing their respective advantages and disadvantages. The latter, deals with surface modification to improve the quality of producing implants and corrosion of Ti implants and also medical aspects such as cell attachment, osseointegration, osteoconduction and biocompatibility and different factors which affect these characteristics as well as leading factors to the failure of implants. The horizon of prospective future work to enhance the durability and quality of producing implants is discussed in the final section.

## 1.2 Ti-Based bio-composites

Tissue attachment to biomaterials is divided into four groups including; nearly inert, porous, bioactive and resorbable. Metal biomaterials fall into "nearly inert" and consequently in recent years work has been conducted to improve biocompatibility and cell attachment bio-composite materials. Table 2 illustrates mechanical properties of materials used for bio-composites (41).

Table 2 Mechanical properties of materials used in bio-composite fabrication

| Property | Bulk HA | TiO2 | ZrO2 | Polyether ether ketone (PEEK) | Polycaprolactone (PCL) |
|---|---|---|---|---|---|
| Density g/cm3 | 3.16 | 4.23 | 6.08 | 1.26-1.41 | 1.09-1.2 |
| Compressive strength Mpa | 500-1000 | NA | 7500 | 80-120 | NA |
| Tensile strength Mpa | 78-196 | NA | 420 | 70-208 | 20.7-34.4 |
| Flexural strength Mpa | 115-200 | NA | 1000 | 3700 | NA |
| Young's modulus Gpa | 11-117 | 230 | 150-200 | 3.9-13 | 0.34 |
| Poisson ratio | 0.27 | 0.27 | 0.30 | 0.38-0.43 | NA |
| Elongation at break % | 3-4 | NA | NA | 1.3-5.0 | 700 |
| Fracture toughness Mpa.m0.5 | 1 | 3.2 | 7-15 | 2.3-2.5 | NA |
| Brinell hardness HBW | 300-700 | 880 | 1000-3000 | 21.7 | NA |
| Knoop Microhardness | 430 | NA | 1200 | NA | NA |
| Coefficient of thermal expansion 10-6.k-1 | NA | 9 | 10-12 | 161-669 | NA |
| Thermal conductivity W/mk | NA | 6.5-12 | 2-2.5 | 0.25-0.92 | NA |
| Tmelting C | 1550 | 1640 | 2400 | 335-343 | 58-63 |
| Tglass | NA | NA | NA | 137-152 | 60 |

NA: Not available,

Among other bio composites hydroxyapatite (HA) coating has been widely investigated because its direct chemical bond with bone that is related to biocompatible mineral component and its synthetic form. HA/Ti-6Al-4V coating is one of the most common bio-composites that provides mechanical strength and toughness while improving the biocompatibility of the produced bio-composites because of its similarity to the chemical composition of bone (42-46). The crystal of HA has a hexagonal structure which is stable in body fluid. Different methods to produce HA is used which will be discussed in the section 9.

## 1.3 Metallurgy of Ti-Based alloys in biomedical applications

Ti alloys are classified into three simple types which contain; α, β and α+β, some elements are dissolved preferentially in α phase such as Zr, Al, Sn, O and Si raising in α+β phase. The addition of these elements results in modulation of the alloy properties, such as hardening and tensile strength improvement. Oxygen plays a dominant role controlling the range of strength for several grades which are called CP-Ti. β phase stabilizes Ti alloys, these are suitable for biomedical application because of their low modulus (which is below that of the α and α+β phase and near human femoral bone) and high specific strength (47).

Some elements stabilize the β phase and depress the α+β phase, these fall into 2 groups: β eutectoid and β isomorphous. Hydrogen molybdenum, tungsten and vanadium stabilize the β phase while oxygen, nitrogen and carbon promote the α phase (22, 48).

Fully α alloys have some limitations in their strength characteristics due to existing reactions that occur at high temperatures notably, in hot forming. These difficulties led to more investigations concerning the α+β phase. This phase contains α with a minimum of 5% β-stabilizing elements, the most commonly used Ti alloy in industry is Ti-6Al-4V which falls in this classification (49, 50). Table 3 illustrates the phases of important Ti-Based biomaterials.

Table 3 Different phases of commercial Ti-Based alloys in biofabrication (16, 34, 51)

| Ti and its alloys | Type of | Ti and its alloys | Type of |
|---|---|---|---|
| CP-Ti-1 | α | Ti-15Zr-4Nb-2Ta- | α+β |
| CP-Ti-2 | α | Ti–5Al–3Mo–4Zr | α+β |
| CP-Ti-3 | α | Ti–15Sn–4Nb–2Ta–0.2 | α+β |
| CP-Ti-4 | α | Ti-13Nb-13Zr | β |
| Ti-3Al-2.5V | α | Ti–29Nb–13Ta–4.6Zr | β |
| Ti-6Al-4V ELI | α+β | Ti-12Mo-6Zr-2Fe | β |
| Ti-6Al-4V | α+β | Ti-15Mo | β |
| Ti-3Al-2.5V | α+β | Ti-15Mo-5Zr-3Al | β |
| Ti-5Al-2.5Fe | α+β | Mo–2.8Nb–0.2Si | β |
| Ti–5Al–1.5B | α+β | Ti–16Nb–10Hf | β |
| Ti-6Al-7Nb | α+β | Ti–15Mo–3Nb | β |
| Ti-6Al-2Nb-1Ta | α+β | Ti–35.3Nb–5.1Ta–7.1Zr | β |

## 1.4 Hardness of Ti-Based alloys in biomedical applications

Hardness and work hardening play prominent roles in biomaterial implants due to increasing resistance against wear and corrosive effects of body fluids. Machining processes such as milling and drilling result in undesirable work hardening in Ti and Ti-Based alloys and although these operations increase the hardness, due to an unwanted mechanism it has a negative effect on the quality of final productions (52-58). Investigations involving implanting Ti alloys in rabbits illustrated that the hardness and fracture toughness of Ti–5Al–2.5Fe and Ti–6Al–4V ELI were not changed before and after 11 months implantation because the microstructure remained unchanged (51). In binary Ti–xTa alloys with increasing Ta content (0–50 wt.% Ta), the microhardness initially decreased, then increased, and finally decreased again as the percentage was increased. In the case of ternary Ti–20Nb–xTa alloys, as a function of increasing Ta content in the range of (0–10 wt.% Ta), the modulus was constant, whereas the microhardness initially decreased and subsequently increased. These variations occurred due to fluctuations of ternary Ti with α+β phase having a higher hardness and lower tensile strength than binary β phase (59, 60). Using laser engineering net-shaping (LEN) for manufacture of Ti–6Al–4V increased the hardness and tensile strengths compared to conventional wrought and cast production methods because a $\alpha'$ hexagonal close-packed (HCP) martensite phase regime was produced, this gave rise to hardness variations ranging from HRC 37 to 57, tensile strengths ranging from 0.9 to 1.45 GPa and breaking elongation from 14% to 11% (61, 62).

The fatigue elongation of Ti-6Al-4V ELI samples examined at various levels showed decreasing trend on high-cycle-fatigue while the tensile strength increased rapidly within the low-cycle-fatigue region. Indeed, the hardness gradient increased from the surface to the core of the samples, and in the next stages of fatigue the internal hardness was equal to the surface hardness. These phenomena occurred because of changing dislocation density in the sub-structures of both near and far from the surface of fatigued samples. However, during the late stages of low-cycle-fatigue, the dislocation density increased rapidly and because it is initially far from the specimen surface then led to decreasing surface hardness (36, 63). Adding 50% zirconium enhanced the hardness of Ti-Based biomaterials about 2.5 fold in comparison with CP-Ti, tensile strength also showed a similar tendency. In this operation two phase structure including a HCP phase and a small amount of bcc phase, caused from swaging above the α+β transition temperature, led to a drastic increasing in the hardness (64, 65). Furthermore, amorphous and glassy alloy ribbons revealed a lineal increase of hardness value on addition of Pd for instance, in comparison to CP-Ti and Ti–6Al–4V alloys which were melt-spun $Ti_{45}Zr_{50-x}Pd_xSi_5$ glassy alloy ribbons illustrated higher hardness and corrosion resistance, good bend ductility and lower Young's modulus (66).

## 1.5  Elastic modulus, fatigue and strain

An elastic modulus is a value that measures a materials resistance to being deformed elastically. This property is important in biomaterial implants especially after surgeries and under loads. Strain measures the deformation of a material under different forces and fatigue illustrates the weakening of a material under periodical loads, these are both important properties in bioimplants due to load deformation during use and the potential fracture and subsequent failure in the short or long term. Materials used in biomedical applications must have a high cycle loading and strength. This very challenging condition is associated with the aggressive in-vivo body environment leads to fatigue failure of metallic, implants (67).

Using Ti/hydroxyapatite as a biomaterial composite caused cracks in the composite surface coated layer, tensile strength was also much lower than that CP-Ti. Modification of this production process led to the fabrication of surfaces without cracks, however because hydroxyapatite was much smoother than Ti the value of tensile strength decreased. Figure 3 (a-b) illustrates that the hydroxyapatite material of a composite surface layer at the interface has a uniform shape with 15% hydroxyapatite volume fraction $V_{ha}$. Increasing the $V_{ha}$ to 22% resulted in a non-uniform distribution and locally stored hydroxyapatite materials which resulted in the bond of composite surface-coated layer to the bulk titanium becoming weak. This phenomenon occurs due to the degradation of the bridging between the bulk titanium and the titanium of the composite surface-coated layer (68).

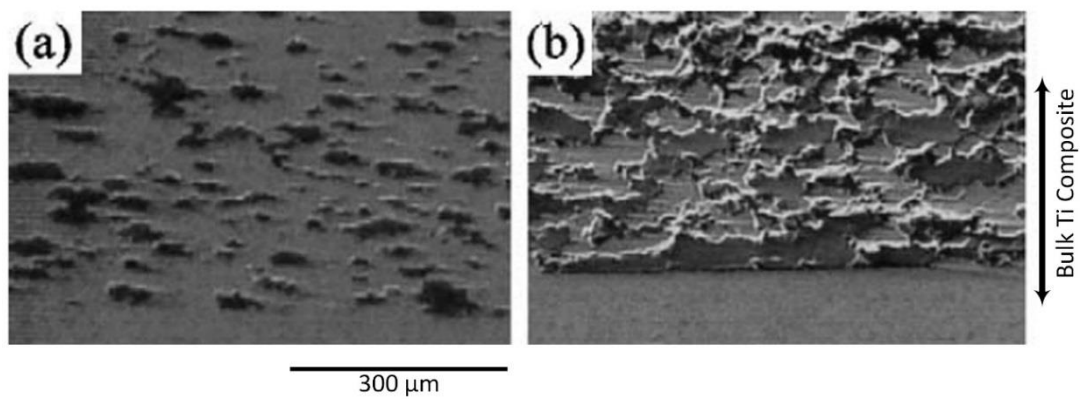

Figure 2 (a) SEM for $V_{ha}$ 15% and (b) $V_{ha}$ 22% coated on Ti (68) (Elsevier licence agreement Number: 3686780699799)

Aging changed the mechanical properties such as elongation until failure and observed microstructure due to phase transformation and deposition of α+β phase leading to increasing tensile strength and decreasing elongation than was observed in the α phase. The tensile strength and brittleness of Ti–29Nb–13Ta–4.6Zr and Ti–16Nb–13Ta–4Mo after aging was equivalent to or greater than CP-Ti alloys, this was related to phase transformation (17, 69). In AM of Ti-6Al-4V an increased breaking elongation of 6.5%-11% was achieved and the fatigue profile was the same as isostatic pressure. In AM annealing improved the breaking elongation due to relieving stresses and refining homogeneous structures (70). Changing the amount of Cr to 3% in Ti-5Nb-xCr due to diffusing of α''+α' phase resulted in modification of the elasticity modulus. β phase appeared by adding 5% Cr and ω phase was diffused by adding 7% Cr which led to increased bending modulus (71). In a quenched binary Ti-Ta microstructure, Young's modulus and tensile properties were related to the Ta content of the alloy. The Ti-Ta alloys showed HCP martensite α' at a value of below 20% Ta, needle-like orthorhombic martensite α'' at 40% Ta, metastable α''+β at above 60% Ta. These changes influenced mechanical properties such as microstructures, Young's modulus and tensile strength (72) (Figure 4).

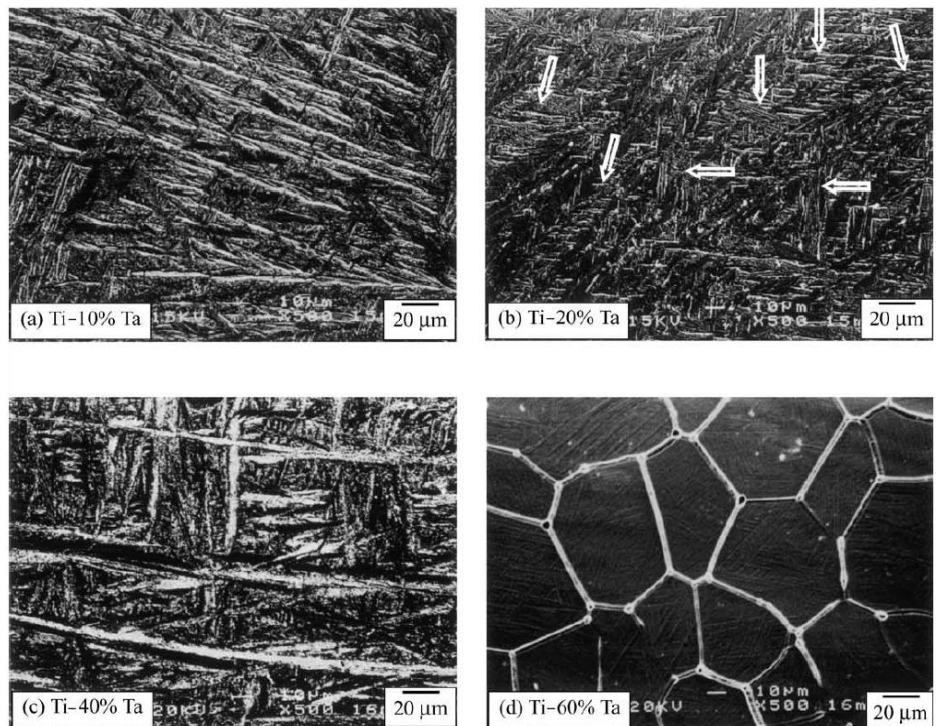

Figure 3 (a) and (b) martensite α' for 10% and 20% Ta with lamellar shape (c) typical martensite α'' with lenticular shape at Ta 40% (d) equiaxed structure for Ta>60% (72) (Elsevier licence agreement Number: 3686780889240)

Changing the amount of Fe in Ti–5Nb–xFe biomaterials had different effects on mechanical properties. Adding 1% Fe caused retention of the metastable β phase while, by adding 4% or more resulted in the β phase being entirely retained in the bcc structure. The ω phase appeared by adding 2%, 3% and 4% Fe. The highest bending modulus was directly related to the formation of this phase, this was observed in 3% Fe while lowest value of ω phase and bending modulus were reported in 2% Fe containing alloys. Furthermore, examination on cleavage facets in the fractured surface showed that by increasing Fe from 2% to 4% (as it can be seen in Figure 5) ductility of the material decreases. The cleavage fractures were highly related to the diffusion of ω phase by adding 3% and 4% Fe, this indicated an extremely low value of bending deflection (73). Machining resulted in changing of the Ti phases, during this phase transformation some intermediate phases such as ω was formed which are brittle and hard to machine, resulting in decreased fatigue life of produced parts (74).

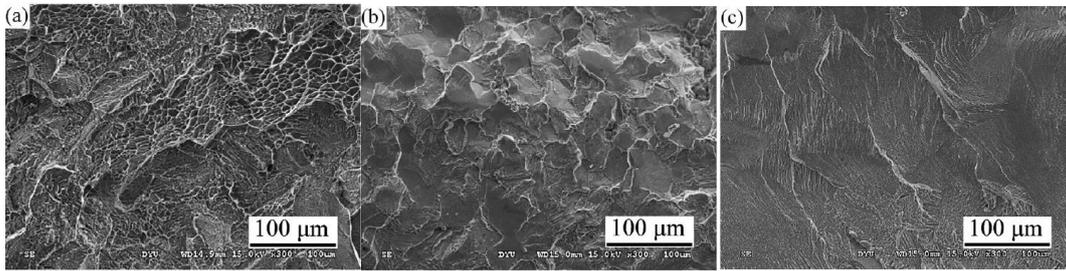

Figure 4 SEM fractographs of Ti–5Nb–2Fe (a), Ti–5Nb–3Fe (b) and Ti–5Nb–4Fe (c) alloys (73) (Elsevier licence agreement Number: 3686781055028)

The introduction of oxygen and nitrogen caused unity in relative growth of Ti and modified it by enhancing microstructure such homogenisation therefore, mechanical properties of Ti-Zr were improved by adding these two elements(75). Ti–24Nb–4Zr–7.6Sn had a low fatigue resistance compared to Ti–6Al–4V ELI because of the effective suppression of micro-plastic deformation by the reversible martensitic transformation and low critical stress that was needed to induce the martensitic transformation. Suppression of isothermal ω phase in cold rolling balanced mechanical properties of Ti–24Nb–4Zr–7.6Sn and decreased Young's modulus while increasing fatigue resistance (76). SEM observations of Ti-6Al-7Nb and CP-Ti proved that a worn surface of Ti–6Al–7Nb alloy was smoother than that of CP-Ti grade 2 and 3.  This was the result of deposition of Al and Nb on the outer surfaces causing softer material, and subsequently a decreased value of hardness (77). Micrometre-sized dendritic β phase deposited in a nano-crystalline matrix such as Ti-Cu-Ni-Sn-M and Ti-10Mo-nNb resulted in a decreased Young's modulus of the composite, while increasing strength and plastic strains when compared to CP-Ti (21, 78). Studies on porous Ti (79) illustrated that the Young's modulus of these materials with approximately 40% porosity was similar to human cortical bone, the rigidity of biomaterials from the least to the most was listed as: cortical bone<Ti<Co-Cr<stainless steels (80). In LEN optimisation of process parameters for production of porous Ti-6Al-4V mechanical properties such as elastic modulus (between 7 and 60 GPa) and the 0.2% proof strength between 471 and 809 MPa were changed. These phenomena occurred due to changes in the porosity and relative density (Figure 6) (81).

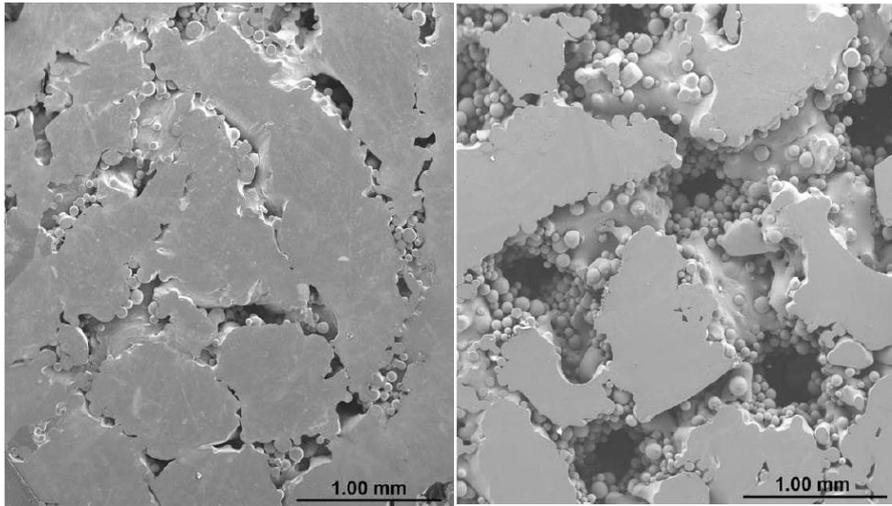

Figure 5 (left) Micrograph of LEN Ti-6Al-4V with 80% relative density (right) with 70 relative density (81) (Elsevier licence agreement Number: 3686781264173)

## 1.6 Surface modifications of Ti-Based biomaterials

The surface characteristics of implants such as surface chemistry, surface roughness, surface potential, surface conductivity and surface energy (hydrophilicity) are very important on initial adhesion, cultivation, and growth of bacteria and subsequent cell action and response. The mentioned characteristics cause protein adhesion and biofilm formation on implants that lead to changes in the biocompatibility and ultimate success of the implant (14, 82). Surface quality plays significant role in secondary processes of different biomaterials such as turning, milling and polishing operation which is subjected for research in recent years (83-87).

The poor tribological property of the Ti-Based biomaterials can lead to increased wear and friction, resulting in a reduction of implant life. Surface modification methods such as coating have been suggested to solve these problems. A combination of surface characterization methods have been recommended for enhancing the quality of the surface from various perspectives and to provide more comprehensive information about the biomaterial surface properties. In order to modify the surface of Ti-Based materials in biomedical applications plasma spray coating, ion implantation, nitriding, carburization and boriding techniques have been employed (88-92). Ti-N and Ti-C-N were produced either by deposition of N and C on the surface with approaches such as physical and chemical vapour deposition (PVD, CVD), plasma nitriding and ion nitriding. This surface modification increases the resistance of biomaterials to wear and corrosion. Optimization of Ti coating processes by using artificial intelligence such as particle swarm optimization and genetic algorithms increased the hardness and resistance to corrosion and wear up to 17% and consequently, this approach resulted in a marked increase in the life of the produced implants (93-96).

Using plasma immersion ion implantation (PIII) and combination of PIII and plasma nitriding (PN) methods improved the surface characteristics of Ti-6Al-4V. Atomic force characterization showed that single step up PIII operation had a higher efficiency than double process because the sputtering in second process removed the implanted layers produced in the first step. The measured values of hardness for both processes increased, and this was confirmed by the nitrogen profile measurement and auger electron spectroscopy (97). Plasma nitriding of Ti-6Al-4V samples enhanced surface characteristics such as hardness and proved that the hardness of nitride layers are highly dependent on the operation time and temperature of these modification processes. Nitrogen diffusion caused permanent lattice strain that resulted in higher surface compressive residual stress, plasma nitrided samples also exhibited lower surface roughness in comparison with un-nitrided samples. Indeed, the research proved that nitrided samples had lower friction forces throughout the fretting cycles at all stress levels (98, 99).

Heat treatment, such as annealing, was used to make unique microstructures, for example bi-modal. This resulted in improved surface characteristics in LEN of Ti–4Al–1.5Mn. Annealing the α+β phase in different temperatures created bi-modal microstructures consisting of coarse crab-claw-like primary α and fine lamellar transformed to the β phase. Figure 7 shows that the fraction area of the crab-claw like primary α drastically reduced with increasing annealing temperatures. The impact toughness of the LEN alloys and wrought productions were the same and both were highly improved by annealing in α+β regions. This improvement was related to the interfaces which were obstacles for crack propagation, contributing to a higher impact toughness (100).

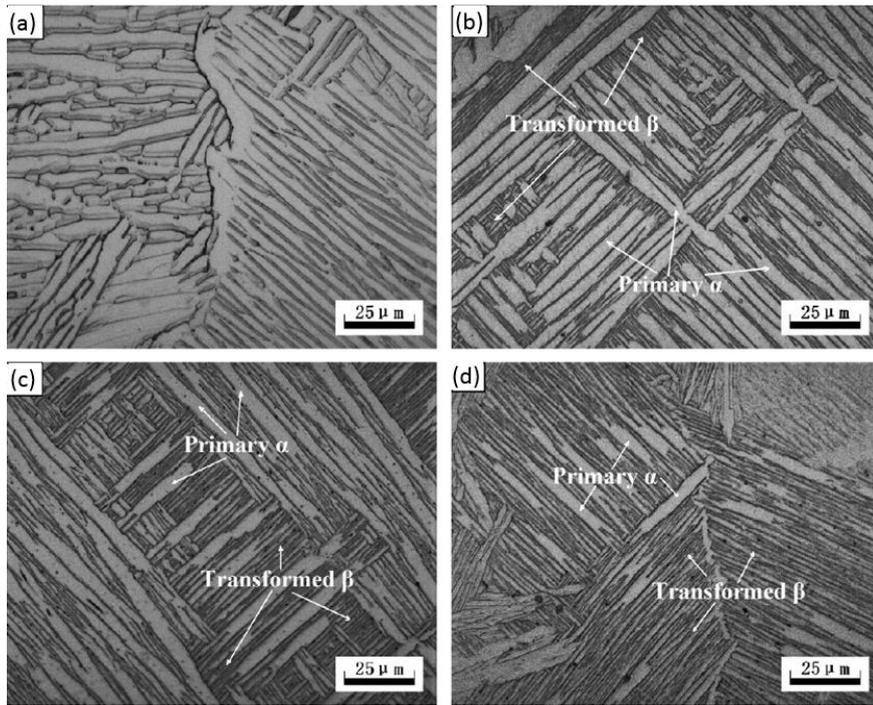

Figure 6 Microstructure of the LEN Ti–4Al–1.5Mn alloy (a) anneal temperature: (b) 945$^0$ C, (c) 955$^0$ C and (d) 965$^0$ C (100) (Elsevier licence agreement Number: 3686781396805)

Laser nitriding was used for enhancing the microstructure of Ti-13Nb-13Zr alloys, including surface roughness and corrosion behaviour. In this process, cracks were not found either on the surface or in the vertical cross-section of the samples nitrided in both $N_2$ and dilute $N_2$+Ar environments due to enrichment of Zr and Ti in the dendrites and improving surface quality. The surface quality was found to be related to the amount of nitrogen, corrosion resistance of the laser nitrided samples in simulated body fluid (SBF) (Ringer's solution) was better (as expected) for diluted samples (101). Another methodology investigated for enhancing surface and mechanical properties of Ti-Based alloys such as Ti-6Al-4V in biomedical applications was low plasticity burnishing (LPB), this procedure was developed as a rapid and inexpensive surface enhancement method. LPB produced a deep layer of compression with nominal cold work of the surface, and could be incorporated into manufacturing processes. Indeed, these layers had high resistance to thermal shocks and overloads and as a result, improved surface characteristics (102). Oxygen diffusion hardening using alpha-tantalum PVD-coatings on titanium improved surface characteristics and frictional properties such as residual compressive stress and resistance to crack. These could be attributed to (a) a decreasing of preferential orientation or an increasing a number of lattice defects that were the results of the incorporation of oxygen atoms into the lattice, (b) oxygen dissolved interstitially in the tantalum lattice, occupying the octahedral sites, (c) the residual compressive stress within the tantalum layer caused by interstitial oxygen or (d) a coherency stresses that led to the perfect lattice matching of tantalum and titanium. Therefore, surface hardness and frictional properties of Ti improved at least 50% and this method could enlarge the field of applications of Ti in orthopaedic implants (103). Indeed, oxidation of the titanium occurred and caused formation of $TiO_2$ on the surface layers. Oxygen diffusion then occurred under these layers and thus, surface features such as hardening and wear were improved (up to 3 times) as illustrated in Figure 8. Formation of oxide layers were accompanied by the dissolution of diffusing oxygen in the metal that was located beneath the surface layer of $TiO_2$ (104, 105).

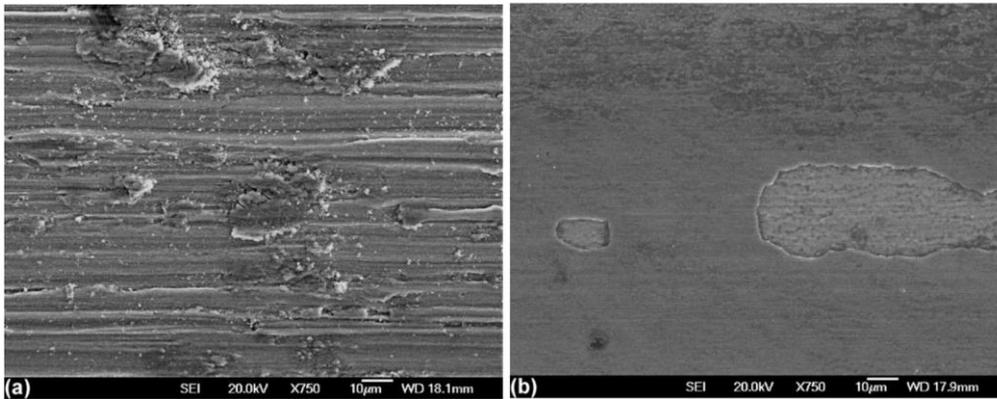

Figure 7 Micrographs of wear impacts for (a) untreated and (b) oxidised alloy after 300 min of testing time (104) (Elsevier licence agreement Number: 3686790034830)

A diamond like coating (DLC) film on CP-Ti formed a protective layer on material surfaces, reducing wear and erosion resulting in increased resistance to deformation of the implants. These layers had the same properties as real diamond, including hardness, chemical stability and wear resistance. Thus DLC could be used as a protectant and lubricant in high abrasion areas, for instance in centrifugal blood pumps or ventricular assist device during heart surgery. Figure 9 shows that the DLC layers under cyclic loading had much higher resistance when compared with those of the non-coated samples. This improvement is related to the hardness of diamond deposited on the CP-Ti surfaces (106, 107).

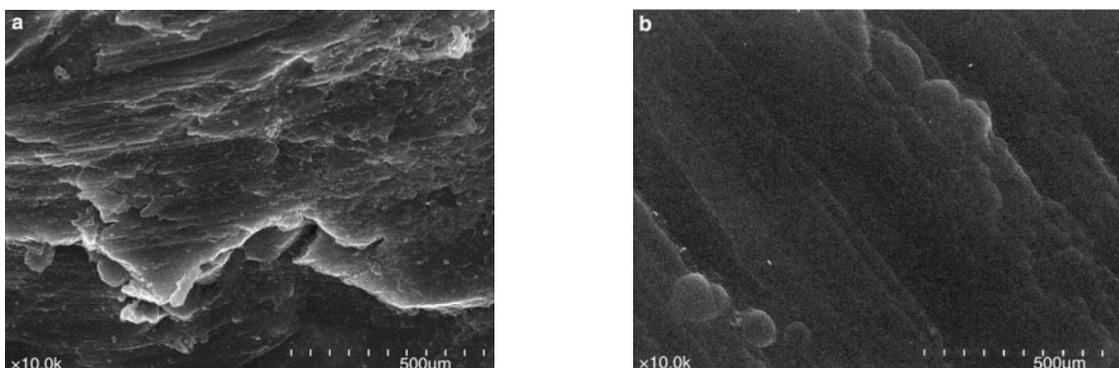

Figure 8 SEM image of titanium implant after cyclic loading (×10 000) (a) a non-coated (b) DLC coated (106) (John Wiley licence agreement Number 3686790284385)

On DLC coated surfaces increasing hydrogen functionalization decreased surface energy or hydrophobicity. It has been shown that this hydrogen value is an important factor in the biological response of DLC surfaces. The reason was attributed to (a) increasing hydrogen caused increasing hydrocarbon bonds. These bonds were representative of oil that was hydrophobic in nature, and (b) "Hydrogen could influence protein adsorption via electrostatic attraction. It is known that hydrogen bonding is the underlying mechanism for hydrophilicity, and the hydrogen atom in the liquid phase must be attached to a relatively electronegative element. The key chemical components of a C:H bonding were C and H. Although both C and H were electronegative, H was less electronegative than C. As a result, surfaces with more H, externally appeared to be a weak array of positive charges compared to those with less H at the molecular contact distance, arising from surface dipoles as polarized covalent bonds". Macrophage cells were spread well on all DLC surfaces, and the surface analysis results showed that the non-toxic nature of the surfaces was enhanced, due to increased cell viability. Also, it was proved that increasing surface roughness and surface energy improved the macrophage cells viability and the albumin: fibrinogen adsorption ratio. (108, 109).

## 1.7 Ti-Based biomaterials corrosion

Although Ti and Ti-Based biomaterials are resistant to corrosion, this characteristic is still important in bio-manufacturing and is currently receiving considerable research attention. The major corrosion problems of Ti are; crevice corrosion, pitting corrosion, stress-corrosion cracking and corrosion fatigue (110-113). Fretting corrosion of different material couples is related to various factors such as normal motion, load and experiment situations (114, 115). Due to the corrosion inhibiting self-healing oxide film ($TiO_2$) Ti is known as a stable metal and has a higher corrosion resistance compared to stainless steel and copper. The Ti surface is sensitive to oxidizing solutions notably, to chloride ions, however, it is resistant to the concentration found in sea water, as well as atmospheric corrosion (22). On the other hand, break down of oxide films due to removal of $TiO_2$ and immersion in highly penetrating corrosive solutions can lead to a drastic decrease in corrosion resistance. For instance, Ti is easily dissolved by hydrofluoric acid, mainly because this acid destroys the $TiO_2$ film on the surface. Generally the α+β and β phases of Ti were observed to possess a high corrosion resistance, due to difficulty in initiating cracks these Ti alloys are highly resistant to stress corrosion cracking (34, 116). Indeed, crevice corrosion occurred in chloride, fluoride, or sulphate solutions at temperatures of $73^0C$ which was higher than the human body temperature and thus will not occur in implants (117, 118).

Increasing Mo concentrations in Ti-Mo biomaterials improved corrosion resistance due to the deposition of metastable α" phase with a fine acicular martensitic morphology. For instance, adding up to 7.5% Mo increased corrosion resistance similar to CP-Ti, while up to 15% was lower than CP-Ti because the amount of Mo in the outer layers of Ti–7.5Mo was smaller than Ti–15Mo (119). Alloys containing Mo, Zr and 0.2% Pd improved corrosion resistance use in biomedical applications due to increasing atom diffusion and subsequently increase in the relative density of near surface layers (22). Micro-abrasion-corrosion tests on the Ti alloy in Hank's solution illustrated that the wear rate was related to the load and corrosion current densities. The reason is a tribo-chemical mechanism which occurs at higher loads and the interaction between micro abrasion, oxide formation and efficiency of oxide removal in different situations. Also, implantation of carbon into Ti–Mo and thermal oxidation led to the formation of Ti and Mo carbides as a protective layers and an increase in corrosion resistance (47, 120-122). In Ti16Nb alloys deposition of Nb around outer surface layers and an increase in relative density led to excellent anti-corrosion properties in Hank's solution, which resulted in an alloy with superior corrosion resistance than that of CP-Ti (123). An anodic polarization test using an automatic potentiostat in 5% HCL solution proved that Ti–Ta was highly resistant against corrosion due to formation of $TiO_2$ passive films which were strengthened by highly stable $Ta_2O_5$ passive films (124). In CP-Ti biomaterials increasing fluoride concentration destroyed $TiO_2$ protective layers, thus the polarization and corrosion resistance decreased (125). Alloying Ti with Ta (Ti60Ta) resulted in the construction of two-layered film structure on the surface and increased corrosion resistance compared to Ti-12Mo and CP-Ti moreover, β alloying elements of Ti improved corrosion behaviour (126).

It has been proven that Ti-6Al-4V has high ferreting corrosion resistance compared to alloying Ti with elements such as Co, Cr, Mo alloys while, CP-Ti has high pitting corrosion resistance rate. Ti-Based alloys in α phase had low resistance to stress corrosion rate, but will crack in a high level of oxygen because of decreasing the hardness. β phase stabilizing elements such as Mo and V improved stress corrosion cracking of Ti due to enhancement of a heat treatment capability and subsequently improving homogeneity and microstructure (127, 128).

Heat treatment improved Ti characteristics against corrosion for instance, quenching in β phase increased anti stress corrosion properties of Ti compared to α+β phase caused by formation acicular microstructure construction. Ageing, quenching and plastic deformation enhanced homogeneity and microstructure in β phase, forming equiaxial grains. This improved Ti stress corrosion resistance. In contrast, wrought Ti production was highly susceptible to accumulation of residual stresses and had low stress corrosion resistance (129). Additionally, producing protective layers by combining Ti-Based and other biomaterials such as Co-Cr, stainless steel and Mg-Based materials resulted in better performance on corrosion fatigue tests. These materials were suitable for dental implants which are normally exposed to high mechanical loads such as pressure, friction and fatigue (130-133). Studies on ultra-fine grained (UFG)-Ti produced by equal channel angular operation in SBF solution illustrated that UFG-Ti had higher corrosion resistance than coarse-grain CP-Ti. The reason being that formation of dense corrosion products and appearing Ca, P and Ti elements probably formed by the interaction between the $TiO_2$ layers and SBF on the surface. Higher corrosion resistance in UFG-Ti was related to the stronger oxide films and quicker passivation on the surface (134).

Corrosion resistance in SBFs for ion implanted surface of Ti–6Al–4V and Ti–6Al–7Nb was enhanced due to the change in the nature and composition of the passive films formed after implantation. Formation of precipitates of TiN and $Ti_2N$ that immobilized underlying titanium atoms, preventing their movement and stabilizing the growth of the oxide film led to these improvements (135).

Controlling anodic oxidation in the production of bio-composites created uniform $TiO_2$ films and improved the bonding strength between HA and Ti substrate. Produced $TiO_2$ was a very good support for HA deposition and increasing corrosion resistance rate on bio-simulated Fusayama-Mayer salvia solution. HA-Ti composite containing 0-10% HA has higher corrosion resistance than CP-TI. Increasing the HA value in this composite reduced corrosion resistance due to making crater-like and local defects induced by ceramic particle detachment (136, 137). Electrolyte deposition of $HA/ZrO_2$ in ZrO $(NO_3)_2$ and subsequent process in the mix of $Ca(NO_3)_2$, $NH_4H_2PO_4$ and NaF on Ti substrate showed that $ZrO_2$ buffer layer improved the bond strength between substrate and fluorine-doped. Double layer coating demonstrated higher corrosion resistance and better mechanical properties that resulted to making dense and uniform nanostructured $F-HA/ZrO_2$ DLC synthesized with electro-deposition and lower dissolution rate.

## 1.8 Biocompatibility of Ti-Based alloys

Biocompatibility is the ability of an artificial material that is used as an implant to perform with an appropriate host response. Thus, clinical interaction of the human body and biomaterial is called biocompatibility (138, 139). Ti was bio-active and bio-inert because of mechanical and chemical bonding with bone. In order to increase the early chemical bonding heat-treatment was suggested (117, 118). New type of Ti β-phase alloys are composed of elements such as Ta, Zr, Nb and Sn. These have achieved good biocompatibility and excellent mechanical properties such as high strength, low Young's modulus and good cold workability and have been more commonly used in recent years (17, 140-142). High value of some elements with β-stabilizing properties in Ti alloys such as Mo are not suitable for biomaterial applications because of possible release to the surrounding tissue. "Two different cells that are aortic endothelial and the osteoblasts on Mo were highly affected from the substratum in their viability (The ability of a living organs or an prosthetic bioimplant to maintain itself or recover its potentialities). The cytoplasm content was totally diminished and cell spreading was reduced on Mo so this element must be used in small value as β stabilizer for Ti-Based biomaterials" (143).

Ni, V and Al in biomaterials were considered to be rather toxic due to ions releasing in the human body while small impurities of $NiTi_2$ and $NiC_x$ with martensite, monoclinic and austenite structure had good biocompatibility (66, 76, 144, 145). Cell culturing on osteoblast cells for Ti–5Nb–xFe alloys after 4 days showed that the rate of cell proliferation was related to the value of Fe and chemical bonding between this element and cells. The cell proliferation level for Ti-5Nb-5Fe was higher than CP-Ti and Ti-5Nb which is shown in Figure 10. The results proved that Ti–5Nb–xFe and Ti–5Nb had a good biocompatibility, viability and support osteoblast cell attachment (73).

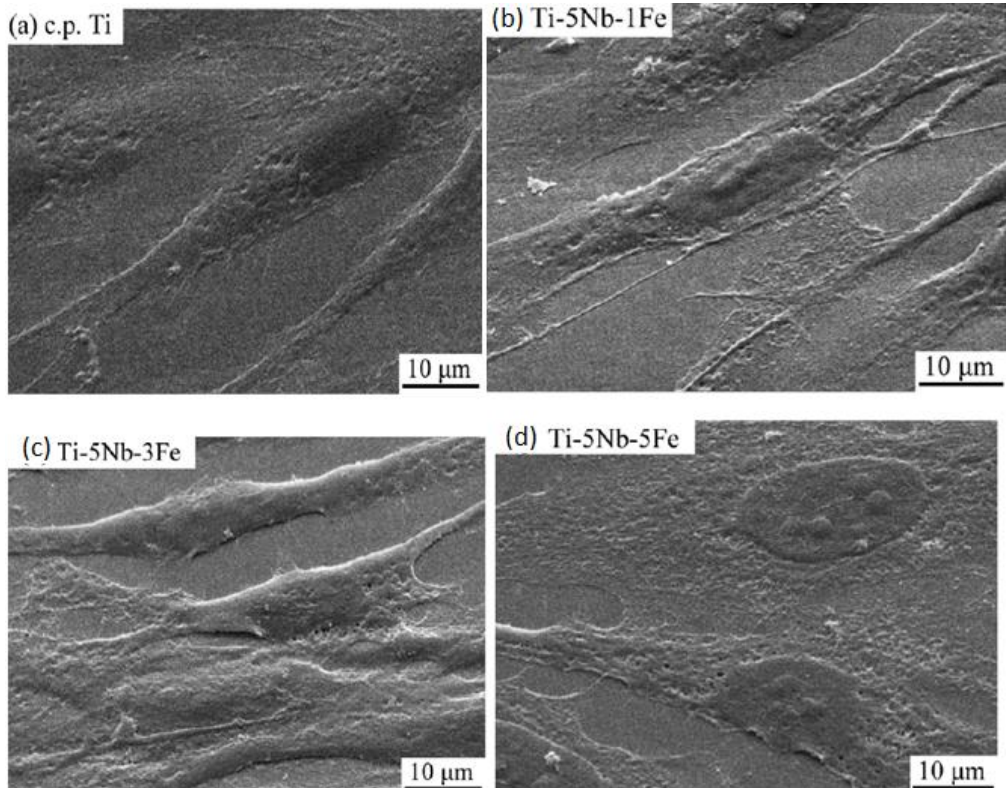

Figure 9 SEM micrographs of osteoblastic cells after 4 days (500X magnification) (73) (Elsevier licence agreement Number 3686790512717)

Evaluation of cytotoxicity for Ti–29Nb–13Ta–4.6Zr on Eagle's culture solution and Zr balls at a temperature of 310K in 7 and 14 days for L929 cells illustrated that cytotoxicity and cell viability for this alloy were the same as CP-Ti that can be related to the β phase and non-toxicity properties of this phase (80). Human osteoblast cell proliferation for CP-Ti, Ti-W and Ti–7.5TiC– 7.5W using micro-culture tetrazolium test illustrated that with increases in incubation time, considerable cell proliferation was observed. This high rate of biocompatibility was attributed to the formation of $TiO_2$ that had good biocompatibility with living tissues regardless of adding TiC and W. No toxicity issues were observed during cytotoxicity tests on L929 mouse fibroblast cells and cell growth had a direct correlation with the time of incubation that is illustrated on Figure 11 (146).

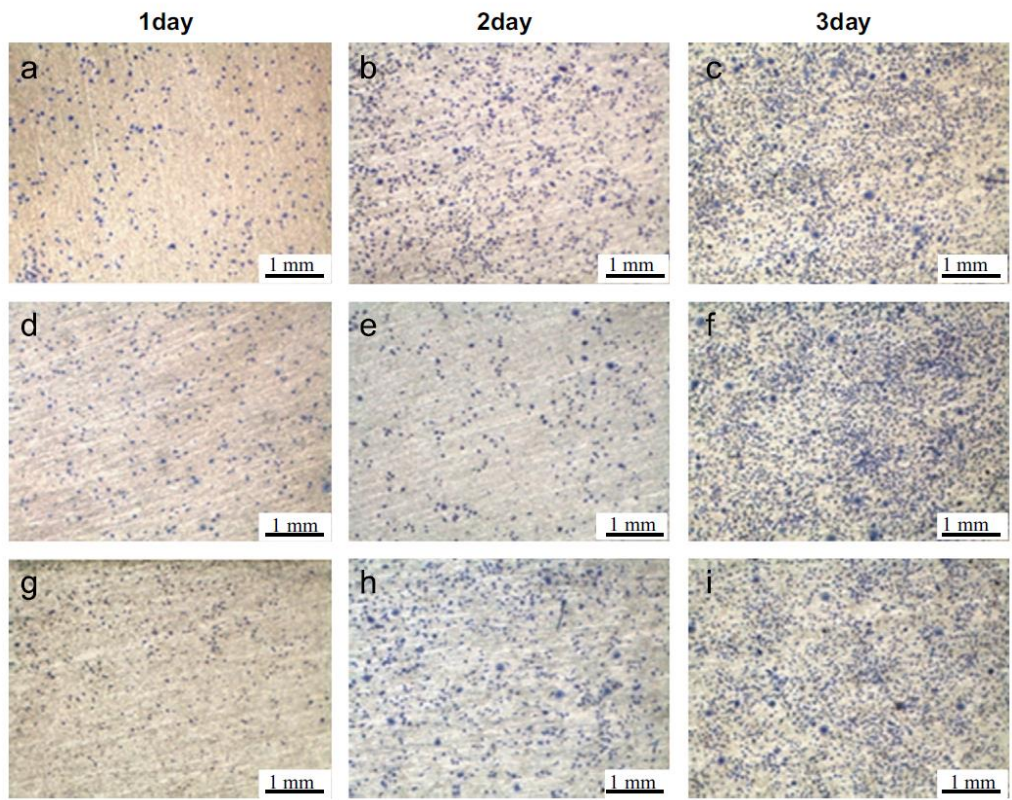

Figure 10 Optical images of L929 cells cultured (a, b, c) CP-Ti, (d, e, f) Ti–10W, and (g, h, i) Ti–7.5TiC–7.5W after being dyed with Giemsa's staining solution (146) (Elsevier licence agreement Number: 3686790659095)

Aging improved the biocompatibility of Ti-Based biomaterials such as Ti-50 mass% by making orthorhombic martensite α''. The formation of this structure was sensitive to temperature and time. In aging various phases such as ω+β, ω+α+β, α+β appeared and the best mechanical properties and biocompatibility were obtained with an α+β phase and temperature of 873K. These properties were related to a lower modulus, and moderate elongation to failure (69). Furthermore, Nb and Ta improved Ti biocompatibility, Ti16Nb and Ti–Ta alloys were not cytotoxic to L929 cells due to an extremely stable oxide layer which was formed on the alloy surface. This layer inhibited the inner metal ion release, shielding the cells. These alloys had an excellent biocompatibility, equal to CP-Ti with a high cell proliferation rate. Figure 12 illustrates cell viability for cells on CP-Ti, Ti16Nb and Ti-Ta alloys after 7 days incubation for solution treatment (ST) and solution treatment and aging plus quenching in ice water (STA) (123, 124).

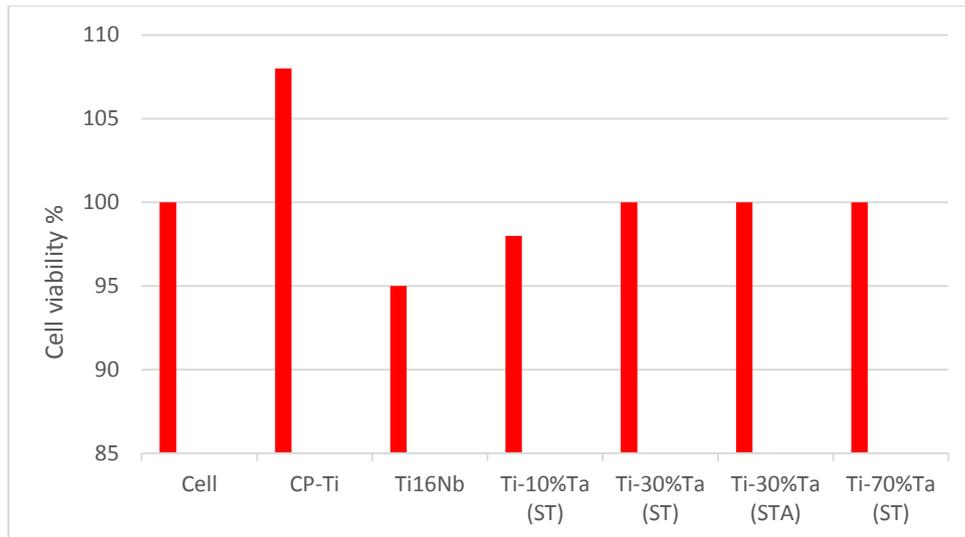

Figure 11 Cytotoxicity test for the TiNb and Ti-Ta alloys on L929 cell after 7 days culturing in extraction mediums (123, 124)

Ti-6Al-4V-xCu alloys had antibacterial characteristics with high corrosion resistance and cytocompatibility, but it is toxic. The antibacterial ability was related to Cu content, which has significant potential for clinical applications as a surgical implant material. Bacterial colonies for two common bacteria in daily life were E.coli and S. aureus, these are shown in Figure 13 after co-culturing on Ti-6Al-4V-xCu and Ti-6Al-4V. It can be seen the number of bacterial colonies after co-culturing with Ti-6Al-4V was significantly higher than Ti-6Al-4V-xCu alloys. Free-form fabrication methods such as electron beam melting resulted in a controlled porosity rate (adjusted by changing process parameters). Materials fabricated in this manner have been found to be ideal for orthopaedic implants due to the effect of changing porosity on biocompatibility (147, 148).

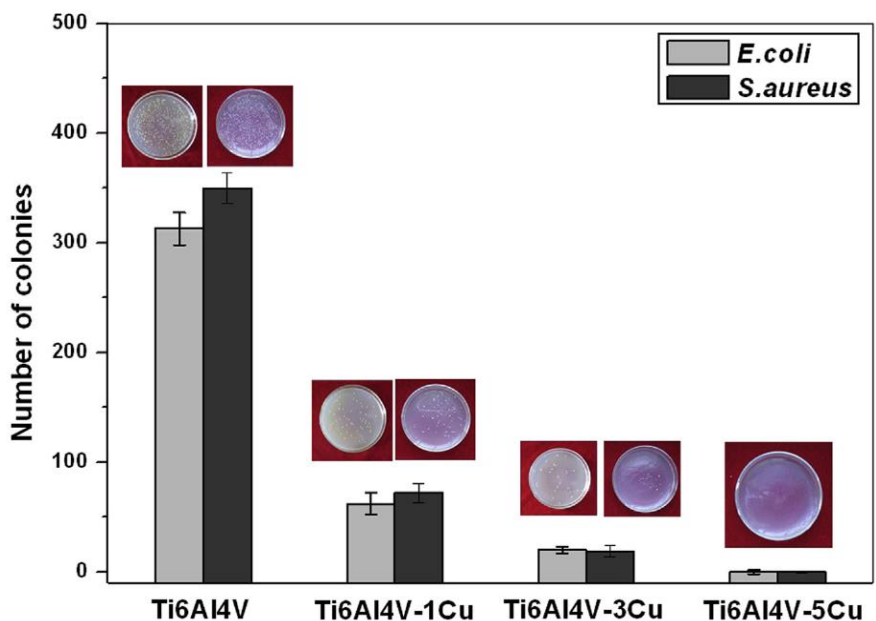

Figure 12 Bacteria colonies value after co-culturing on Ti-6Al-4V and Ti-6Al-4V-xCu (147) (Elsevier licence agreement Number: 3686790804763)

Investigations of 8 weeks duration on implanted biomaterial containing graded layers mixed by Ti/hydroxyapatite (HA and Ti) proved that tissue reaction occurred gradiently in response to the graded structure. Small amounts of decomposed products of HA phase (called biodegradable α-TCP and $Ca_4O(PO_4)_2$) appeared in the graded layers mixed by HA and Ti. Other compositions were not observed in the fabrication process. Indeed, it was found that newborn bones between Ti/hydroxyapatite and host bones grew actively and had a growing behaviour from the edge of host bones to the implants which demonstrated that no important defensive reaction occurred between implants and body tissues. This result is probably related to HA compatibility with living tissue, confirming the high biocompatibility of this material (149, 150).

$TiO_2$ particles on the surface of Ti-Based alloys with diameter of 50–90 nm improved biocompatibility, this is thought to be attributed to high homogenisation and anti-corrosion characteristics hence, oxidation of Ti is a commonly used technique in bio-manufacturing (151). Cytotoxicity of elements in Ti-Based biomaterials from the most to least are listed as; Cu > Al=Ni > Ag > V > Mn > Cr > Zr > Nb > Mo >Ta>Sn> CP-Ti.

Generally, Ti-Based biomaterials had higher than 80% cell viability, for instance, Ti–10Nb alloy exhibited the highest cell viability (124.8%), which was higher than that of CP-Ti. The ranking of cell viability for pure biomaterial ingots from the most potent to least potent is shown in Figure 14 (152).

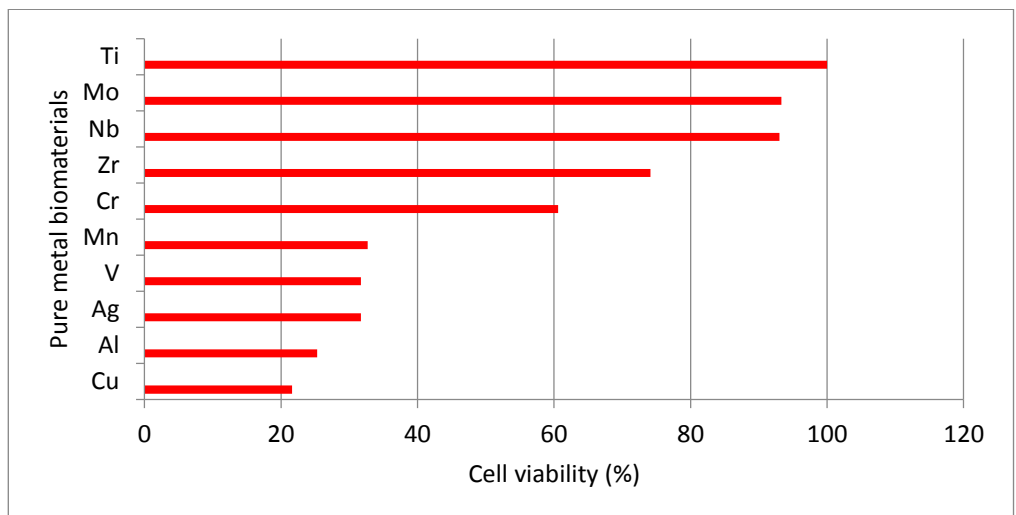

Figure 13 The cell viability for pure metal biomaterials (152)

Nano composite coating HA/calcium silicate was reported as having a porous structure that resulted in enhanced cell attachment and osseointegration. Strength tests in this nano-composite illustrated that HA/CaSiO₃ had a higher bond strength than HA-Ti composites. Moreover, the proliferation of MC3T3-E1 osteoblast cells on HA/CaSiO₃ had higher rate compared to HA-Ti bio-composites.

HA/CaSiO$_3$ had higher stability in physiological environment and better corrosion resistance. The mentioned improvements are related to the porous structure of HA/CaSiO$_3$ and resemblance to human natural bone that made it compatible with the human body.

Modification of process parameters in producing HA-Ti bio-composite using plasma spraying to obtain optimum thickness and surface coverage has been carried out by Huang et al (153). The study demonstrated that the crystallinity level improved after immersing in the SBF and, crucially, no calcium ion release of vanadium was observed that resulted in high biocompatibility for bone generation. Uniform surface coverage and thickness in the range of 47—130 μm was achieved after modification of nozzle transverse speed and Ti surface rotational speed. Indeed, increasing the amount of hydrogen or decreasing the powder feeding rate melted HA particles completely that led to higher adhesion strength, denser and uniform bio-composites. Sol-gel disillusion to produce three bio-composites including HA, fluor-apatite (FA) and fluor-hydroxyapatite (FHA) to understand the bond strength and the interaction of these materials on Ti substrates was investigated by Tredwin et al (154). The study demonstrated that all three materials offered superior alternative for coating Ti bioimplants. Coating thickness had direct and indirect correlation with increasing fluoride ion substitution and spin coating speed respectively. Indeed, increasing fluoride ion substitution and heating temperature resulted in increasing bond strength and subsequently increasing stability, decreasing micro-motion and finally improving biocompatibility.

## 1.9 Osseointegration, osteoinduction and osteoconduction for Ti-Based alloys

Osteoinduction is related to the bone healing process and shows the recruitment of immature cells to develop into preosteoblasts. In bone healing, issues such as a cracks and fractures are highly associated with osteoinduction. Osseointegration is a direct structural and functional connection between living bone and the surface of a load-bearing artificial implant and is directly related to mechanical stability. Osseointegration decreases as a function of increasing micro-motion of implants, blood vessel growth and fibrin adhesion. The growth of bone on the surface of the implants is osteoconduction and is related to osseointegration. Improving these three characteristics lead to enhancing the quality of the implanting process and decreasing implant defections, improving patient outcomes. Cell morphology orientation, attachment and growth are highly related to the quality of surface. Some chemical and biological reactions occur after implantation of biomaterials such as adsorption of water and proteins. These are related to the surface properties of the material such as surface chemistry, surface topography, surface roughness and energy. Figure 15 shows one of the following procedures will happen after implanting process in human body (9, 155).

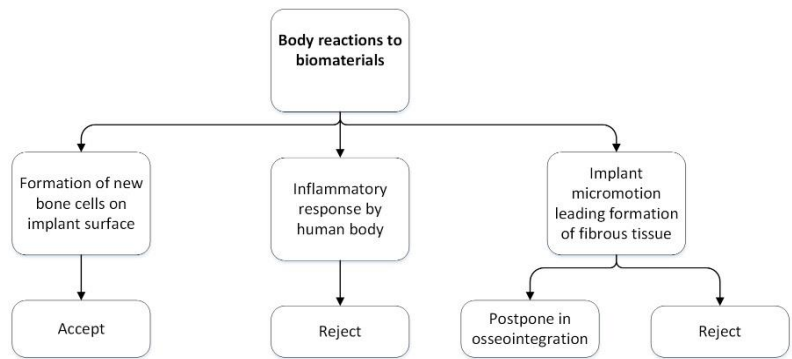

Figure 14 body reaction to biomaterials after implantation (9)

Adhesion is required for embryogenesis, wound healing, immune response and biomaterial tissue integration. Proteins are involved in adhesion to extracellular matrix (ECM) proteins, cytoskeletal proteins and membrane receptors. Interaction of these proteins induced signal transduction and therefore led to cell growth. Cell shape and cyto-skeleton alignment was attributed to the surface topography of grooved surfaces because regular slots aligned cells and increased adhesion. Grooved surfaces, enhanced osteoblastic cell adhesion, attachment and proliferation more than rough surfaces (156). Anodic oxidation caused formation of porous oxide films, these porous structures increased the frictional forces between the implants and surrounding tissues therefore increased osseointegration and biocompatibility. Micro-pores on the material surface increased the surface roughness generated during anodic oxidation while samples with smoother surfaces were more likely to form thicker porous encapsulation. Rough surfaces produced by treatment of surfaces containing micro-pores had a positive effect on the bonding of implants and tissues with the material. Ca and P enriched oxide films were found to have a low amount of micro-cracks, this property improved surface characteristics such as intermediate roughness and crystallinity. In-vitro experiments illustrated that pre-osteoblast cell growth and metabolic activity on Ti and porous Ti scaffolds were comparable. The dispensing angle and size of the powder were found to be important factors governing the final architectural and mechanical properties of the Ti scaffolds (157, 158).

Thermal and chemical improvements of titanium surfaces resulted in changes to the surface topography, oxide chemistry, wettability and protein/cell-binding affinities. Heating, either in atmosphere or pure oxygen, led to an enrichment of Al and V in the surface oxide. Subsequent heating in peroxide solution and exposure to oxygen/atmosphere followed by butanol rinsing decreased the value of V, however no significant change was observed for Al content. This process resulted in a thicker oxide layer and a more hydrophilic surface compared to passivated controls. Heat treatment in normal atmosphere or pure oxygen increased the amount of $Al_2O_3$ on the surface, this increased fibronectin-promoted cell attachment. This treatment in atmosphere with or without a butanol treatment step improved protein induced cell-adhesion, while in oxygen didn't have significant effect. Generally, heat treatment had no inhibitory effect on basal MG63 cell attachment. Protein-induced attachment of MG63 cells to the implants fluctuated with changing V and Al surface composition. The reason of these phenomena were highly reliant on change in surface chemical composition (reactions of the metallic surfaces, caused by reduction in V content) and decrease in the V/Al ratio (159). Staphylococcus aureus (S. aureus) adherence to the ECM and plasma proteins that were deposited on biomaterials was an important issue in the pathogenesis of implant related infections as these are a major cause of medical problems after implantation surgeries. Immediately after implanting biomaterials in the body they become coated with host plasma constituents, including ECM which can be detrimental to the success of the surgery. "Poly (L-lysine)-grafted-poly (ethylene glycol) (PLL-g-PEG) adsorbed from aqueous solution on to metal oxide surfaces, effectively reduced the degree of non-specific adsorption of blood and ECM proteins, and decreased the adhesion of fibroblastic and osteoblastic cells to the coated surfaces". Coating Ti surfaces with any type of copolymers rapidly reduced the adhesion of S. aureus to the surfaces, this is shown in Figure 16 for smooth Ti ($Ti_S$) and rough Ti ($Ti_R$). More bacteria are seen on the uncoated surfaces (a–b) in comparison with coated surfaces (c–d). The reason for this change is that S. aureus attached to the Ti surfaces and can be seen forming clumps of cells, while on the coated surfaces gaps and holes that were produced by copolymer components, decreased the adhesion rate (160).

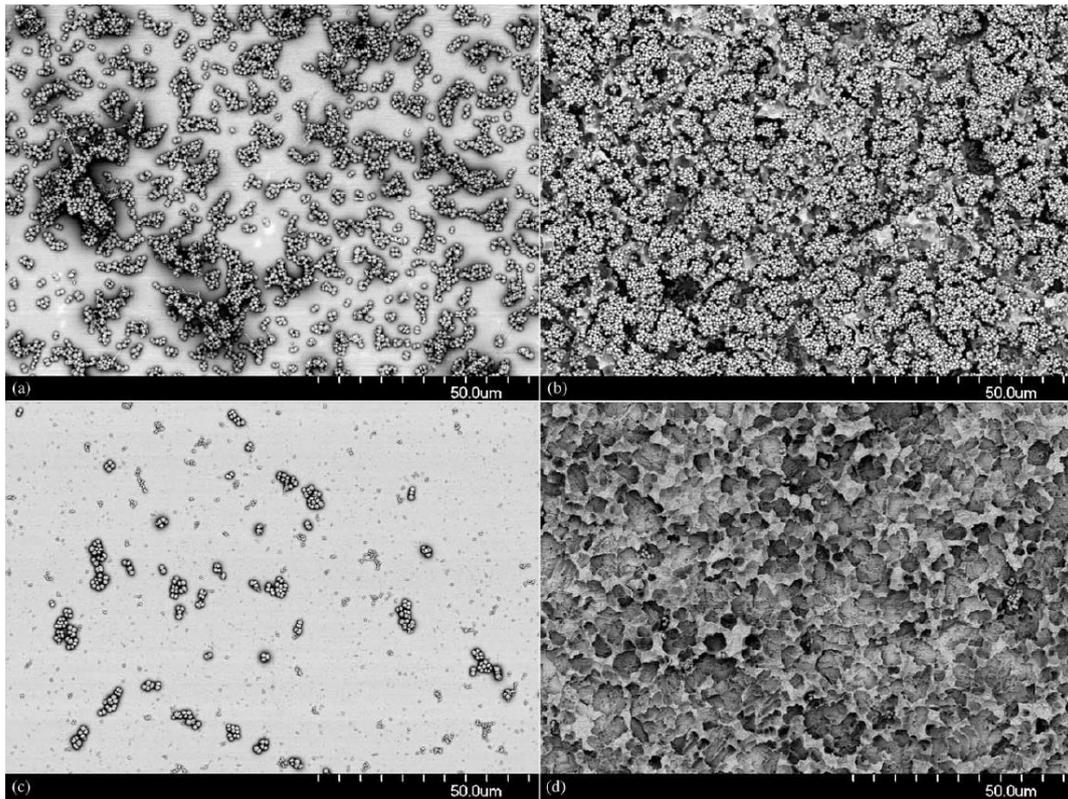

Figure 15 BSE images of S. aureus cultured on the different surfaces for 2 h at 37$^0$C: (a) Ti$_S$, (b) Ti$_R$, (c) Ti$_S$-PEG, and (d) Ti$_R$–PEG (160) (Elsevier licence agreement Number: 3686790993889)

HA coating of Ti-Based biomaterials using different processes such as plasma spraying, sol-gel process and biomimetic growth is one of the most promising methods for promoting osseointegration and osteoconduction. Cells in the HA/TiO$_2$ double layer are exposed to a uniformly dense, homogeneous, well-crystallized structure with high corrosion resistance. This resulted in superior bone integration optimising oxide thickness and good interfacial adhesion and subsequently, a higher degree of osseointegration and biocompatibility than that of the TiO$_2$ single layer coated and CP-Ti surfaces (161-166). In biomimetic apatite coatings formed on micro-arc oxidized titania low voltages caused porous microstructure with completely spherical pores and homogeneous distribution. As a function of increasing voltage the pore size increased and cracks with irregular and rough surfaces appeared. In this process oxide films initially released and dissolved Ca and P and likewise the formation of apatite on the surface of Ti in SBF was highly related to the amounts of Ca and P present. After 14 days Ca and P containing precipitates appeared on the surface of the oxide films and reduced on SBF. The proportion of coated appetite increased and successively bone-like apatite (bioactivity and biocompatibility) were improved (167). Porous HA/collagen composite biomaterial has osteoconductivity and was able to act as a scaffold in formation of bone. This composite has bone conductive activity and was able to unite with bone. The bonding has been related to the similarity with natural bone and inducing the development of osteogenic cells and bone-remodeling units (168). Ti coated with HA, albumin-apatite or laminin-apetite produced by immersion of NaOH and heat treatment in calcium phosphate solution were investigated by Uchida et al. (169) to analyse activation and adhesion of platelets. The results showed higher platelet adhesion and activation for heat-treated samples that can be related to thromboresistance nature of these composites superior that CP-Ti.

Nano-composite with a grain size of less than 100 nm, due to increasing consolidation between HA and Ti substrate, improved various properties of the composite such as hardness value, young's modulus and corrosion resistance. Nano-composites also increased tissue growth cell activity and cell adhesion for osteoblasts because osteoblasts tended to adhere at particle boundaries and nanophase metals have a higher percentage of particle boundaries on the surface in comparison with wrought materials (170, 171).

A mixture of HA with an alkaline dopamine solution deposited on Ti surfaces improved adhesion, proliferation and mineralization of osteoblasts. Also, this process immobilized HA nano particles helping to increase osseointegration. This enhancement can be related to (a) low process temperature, which led to avoiding the damage to HA crystallinity and (b) aqueous process condition and low chemical reactions and damage (172, 173). Investigation on nano-HA-Ti and nano-HA/collagen-Ti on osseointegration onto bone surface proved that nano-HA/collagen surrounded by new bone tissue without encapsulation of fibrous. HA/collagen had great potential for bone contact and bond strength to the bone and nano-HA displayed preferential accumulation proximal to the cell membrane that can be related to nano-grain structure as natural bone. These investigations demonstrated that the application of nano technology on production of HA-Ti composites had good potential on osseointegration, cell culturing rate and grow (174, 175).

Protecting and polishing the surface of biomaterials such as carbon fibre-reinforced composites (CFRC) by carbon-Ti coating operation significantly enhanced the biocompatibility in terms of lower release of carbon particles and higher colonization with MG63 and vascular smooth muscle cells. Combining these two surface modifications resulted in large improvements to the biocompatibility of CFRC, probably because of the development of a biocompatible lattice for assembly of osseous and vascular tissue that could functionally replace a living bone (176). Pulsed direct-current plasma enhanced chemical vapour deposition of DLC and polycrystalline/amorphous $TiO_x$ (DLC-$TiO_x$, $x \leq 2$) and DLC-$SiO_x$ showed no evidence of Ti-C bonds on the surface, Si and C bonds were observed to form siloxane structures. The cell count compressive stress and hardness of the films increased with decreasing Ti content. TiO$_2$ on the surface seemed biocompatible and the cell morphologies on all DLC-TiO$_x$ surfaces appeared conductive to healthy proliferation. Osteoblast cell adhesion in modified DLC films increased with deposition of TiO$_2$ due to increasing hydrophilicity and surface energy in DLC-TiO$_x$ surfaces compared to pure DLC. This led to enhanced in osteoblast proliferation. However, osteoblast proliferation properties were unchanged by the deposition of SiO$_x$ on the DLC films. Figure 17 shows multiple microvilli and spherical structures on the surface that provide continuous exchange between the environment and the cell surface. Large lamellipodes indicated homogeneous colonisation, cells on DLC-TiO$_x$ appeared to exhibit more flattening of the substratum. Furthermore, increased size of cytoplasmic extensions in various directions illustrated excellent adhesion osteoconduction and osseointegration (177).

Oxidation treatment and alkali treatment on Ti–29Nb–13Ta–4.6Zr led to the formation of a titanate layer on the preoxidized surface, and growth of a layer for Ca–P after immersing in SBF or fast calcification solution. This phenomenon resulted in making hard and bioconductive surface and improvement of its wear resistance, bioconductivity and bioactivity (178).

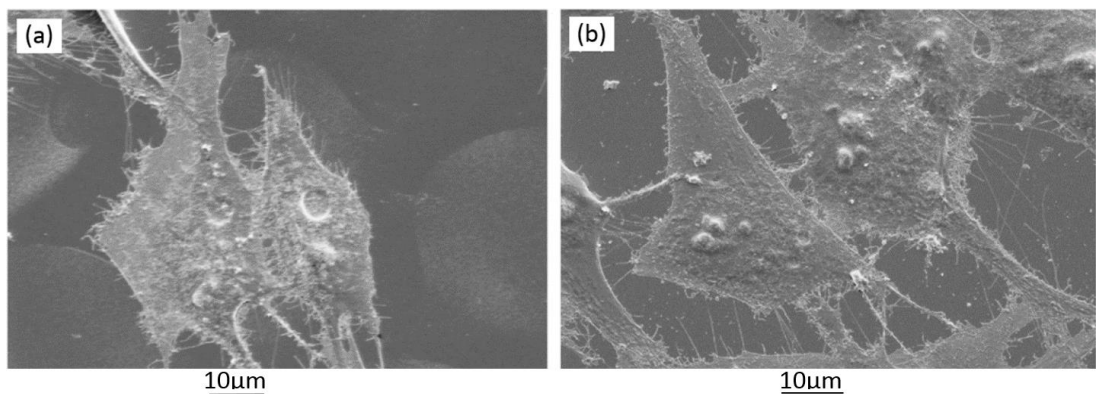

Figure 16 SEM images for osteoblasts growing on DLC (a) and on DLC-TiO$_x$ (b) films (177) (Elsevier licence agreement Number: 3686791131615)

The effect of surface finish (pore size) on the osseointegration of laser-treated surfaces showed that surface blasting considerably improved the osseointegration of laser-textured Ti-6Al-4V implants. Surface blasting of laser-textured Ti-6Al-4V implants with 200µm pores shown the highest level of osseointegration because smaller pores led to decreasing mechanical stability while larger pores showed slower bone implant contact and osseointegration. Enhanced biomechanical stability and higher resistance to fatigue loading on surfaces with 200µm pores was attributed to bone ingrowth through the pores that led to interlocking of the surrounding bone tissue with the implant (179). Figure 18 illustrates different methods that have been developed to improve biomechanical compatibility, cell growth and fixation.

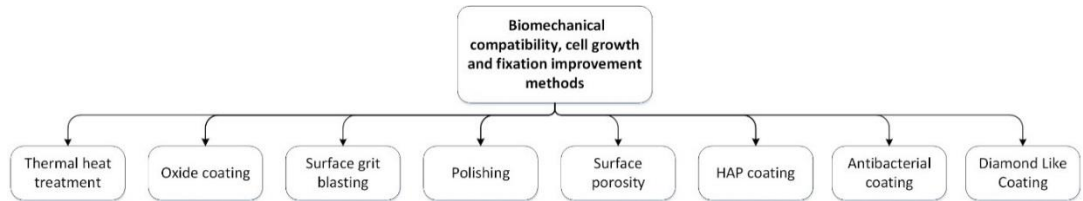

Figure 17 Common methods for enhancing biocompatibility of Ti-Based biomaterials (82, 157-161, 166, 167, 176, 180)

## 1.10 Fabrication of Ti-Based biomaterials

In this step Ti fabrication methods in commercialized systems for biomedical application with determining their advantages and limitations are discussed.

### 1.10.1 Casting and Powder metallurgy

Casting is a low cost method that is used to produce the net shape of raw biomaterials including Ti and Ti-Based alloys (181-184). Improving net shape casting technology, fatigue properties and decreasing metal mold reactions lead to increasing the quality of casting productions (34, 185-187). Ti casting is carried out by two methods including conventional and investment methods. In conventional methods mold material is formed from rammed graphite that produce complicate shapes, has a good surface finish after polishing and minimal reaction rates (188, 189). Another Ti casting method is investment casting which uses a wax mold and is a lower cost method of producing intricate and net shape productions.

This method results in more surface defects, lower surface quality and greater dimensional deviations than graphite mold casting (190-192). Figure 19 shows the surface profile for Ti-6Al-4V and CP-Ti produced using an investment casting process and 3D printing (SLM). The surface roughness was measured in the range of 2.3μm. These surfaces are of inferior quality and not suitable for direct use and would need machining or polishing operations. But the roughness in this operation is better than SLM because in SLM remained particles decrease produced surfaces in the range of 19μm.

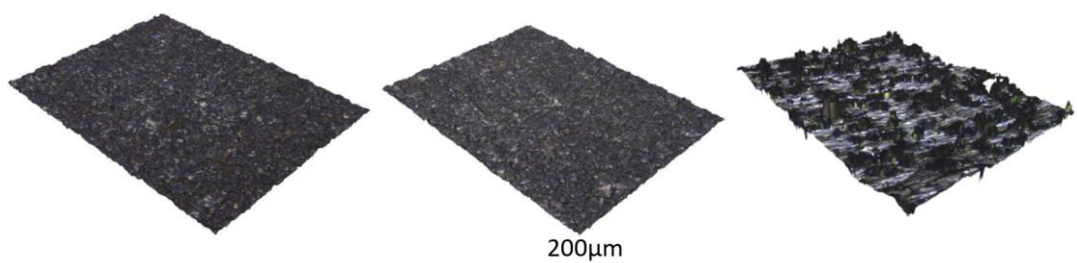

200μm

Figure 18 Casted surface for CP-Ti (Left) Ti-6Al-4V (Middle) and 3D printed (SLM) surface of Ti-6Al-4V (Right) (193)

Powder metallurgy is used for making Ti productions in medical applications which are close to final size (near net shape), resulting in reduced machining operations and fabrication costs. Direct gas atomization, blended elemental technique, rotating electrode and metal hydride reduction are the common methods for producing Ti powder. Among these operations metal hydride reduction and blended elemental technique result in a higher density and are more common for Ti implant fabrication. Figure 20 shows the level of density in common Ti powder production methods. A new approach to powder production is metal hydride reduction. In this method Ti is produced from Ti dioxide in a chemical reaction at a temperature of $1100^{0}C$ to $1200^{0}C$ (below melting temperature). The chemical reaction is shown in Equation 1.

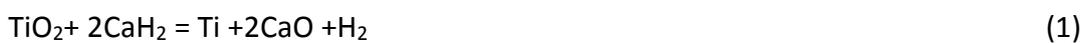

$TiO_2$+ 2CaH$_2$ = Ti +2CaO +H$_2$                                                 (1)

Ti produced from this method has a small amount of chloride and large levels of hydrogen present which can be removed by an annealing process (22, 194). The second common process for making Ti powder is the blended elemental technique (elemental method) in which Ti particles are blended in a twin cone blender at room temperature and a high pressure of 400Mpa (195-197). For obtaining close to 100% density hot pressing, sintering and hot isotactic pressing is suggested. Sintering is performed in β phase and hot isotactic pressing is processed in α+β phase.

Generally, powder metallurgy is used for the forming of complex shapes/composites with uniform microstructure and requires a few or no secondary operations, making it cost and time efficient. Dimensional deviations are low and tolerances are quite high in this method. A high production rate is another advantage of this method. Ti-Based alloys can be produced with infiltration and impregnation of other materials with different physical and mechanical properties such as hardness, strength, density and porosity that have compatibility with human organs with low scrap rate.

This method has limitations on size and dimensions of the productions, especially in hot pressing based techniques. Furthermore, producing powder mold and compression equipment such as pistons are expensive and needs exclusive design for each specimen. Low ductility, strength and fire hazard for Ti and Ti-Based alloys due to low thermal conductivity as well as health problems are another disadvantages of this method.

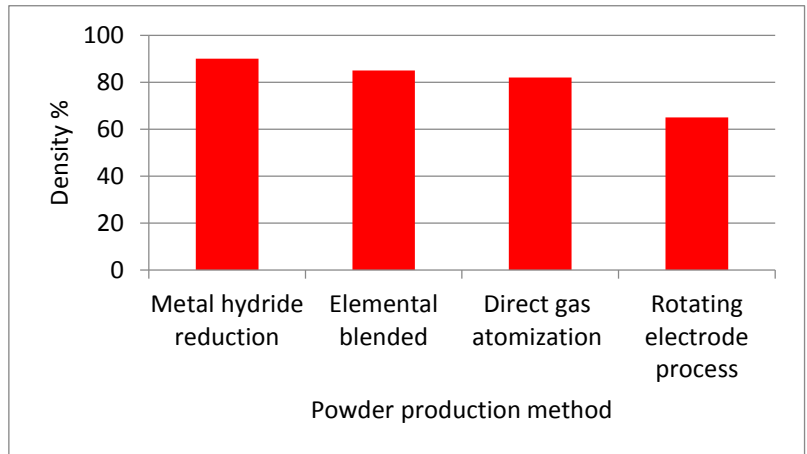
Figure 19 Ti powder making methods (195)

## 1.10.2 Cold working and hot working

Ti alloys have low capacity to be cold worked due to behaviours such as shape memory and spring back, properties related to low modulus and high strength. Strain hardening, expensive equipment, undesirable residual stresses and less ductility are other disadvantages of Ti cold working. For increasing the cold working capacity of Ti hot sizing and stress relieving are recommended. Advantages of cold working of Ti include good control of dimensional deviations, better reproducibility, improved strength, as well as high levels of straightness and machinability (198-203).

Hot working of Ti reduces yield strength, so less energy and force are needed. It is also easier to perform and results in increases in ductility and removal or reduction of chemical in-homogeneities due to the elevated temperature and diffusion involved. Indeed, the size of pores may decrease or close completely during deformation. Controlling this size has a positive effect on cell adhesion, biocompatibility and osseointegration. Hot working of Ti is performed at a slow rate thus increases the production time and cost. Shape memory, workpiece and tool oxidation and lubrication problems can all have a negative effect on the quality of produced materials. Undesirable reactions between the metal and the mold/surrounding atmosphere, poor tolerances due to warping from uneven cooling, thermal shock and variety in grain structures are other problems of hot working techniques. In the temperature ranges above $550^0$C due to oxygen absorption and low thermal conductivity defects can appear on the surface (204, 205). Hydrogen that is absorbed during the hot working process results in poorer mechanical properties, a preheat at a temperature of $200^0$ to $250^0$ C is recommended to increase the efficiency (22).

### 1.10.3   Machining and laser forming (additive manufacturing)

Ti and its alloys are well known to possess a low rate of machinability due to their hardness and low thermal conductivity compared to other metal and ceramic biomaterials. Deformation mechanisms during machining of Ti alloys is a complex process-abrasion, attrition, diffusion–dissolution, thermal cracking and plastic deformation are the main tool wear mechanisms (206-209). For improving the machinability of Ti in biomedical applications it is recommended to use ultra-hard or super hard coated cutting tools such as cubic boron nitride or diamond carbon coatings (210-215). Coated cutting tools such as $TiCN/Al_2O_3$ increase the machinability of Ti and improves the resulting surface characteristics (216). Free Ti machining leads to an increase in tool life and decrease in cutting temperature; however has a negative effect on ductility and impact resistance. High coolant pressure increases tool life and machining efficiency (217-221). In machining of Ti and Ti-Based biomaterial the temperature can reach $300^0C$, thus it is difficult to achieve cutting speeds of over 60m/min (221). Extreme pressure, mineral oil, chemical or synthetic coolant fluid is recommended for the machining of Ti-Based biomaterial in order to decrease cutting temperature and tool wear. Furthermore, in the dry machining of this material localized flank wear is a significant cause of tool failure-brittle fracture of the cutting edges that is observed (222). Machining results in changes of Ti surface characteristics such as roughness, patterns, wettability, surface mobility, chemical composition, electrical charge, crystallinity, modulus and heterogeneity to biological reaction that are important in cell adhesion, osseointegration osteoconduction and biocompatibility. Production of intricate shapes especially by using 5 axis machining, good surface finish, high accuracy in terms of dimensional deviations and selectable surface roughness and subsequently different cell viability, cell growth, osseointegration and biocompatibility by changing cutting conditions and surface topography are the advantages of biomaterials machining. The only disadvantage of this method is expensive machine centre and equipment (223-225).

Additive manufacturing is a process for producing functional prototype parts directly from computer models. This is called additive layer manufacturing and is achieved by deposition of powdered material in layers and the selective binding of the powder using ink-jet printing to produce the net shape components (226-228). Complex and expensive (near net shape) Ti-Based biomaterials are manufactured using this method (229, 230). Different laser processing methods such as selective heat sintering (SHS), selective laser melting (SLM), selective laser sintering (SLS), electron beam melting (EBM) and 3-D laser cladding are used in additive techniques to fuse (deposit) Ti powder at a desired location. The operation is controlled by computer numerical control and the size of powder particle is in micro scale. Figure 21 (Left) illustrates prosthetic acetabular hip designed by SolidWorks software and (Right) shows specimen manufactured by SLM method (Diameter 69mm). In laser forming the quality of produced parts is associated with the dimension of the laser focus, scanning speed, power rating of the laser, size of the powder particles, layer thickness, process atmosphere situations and track overlap.

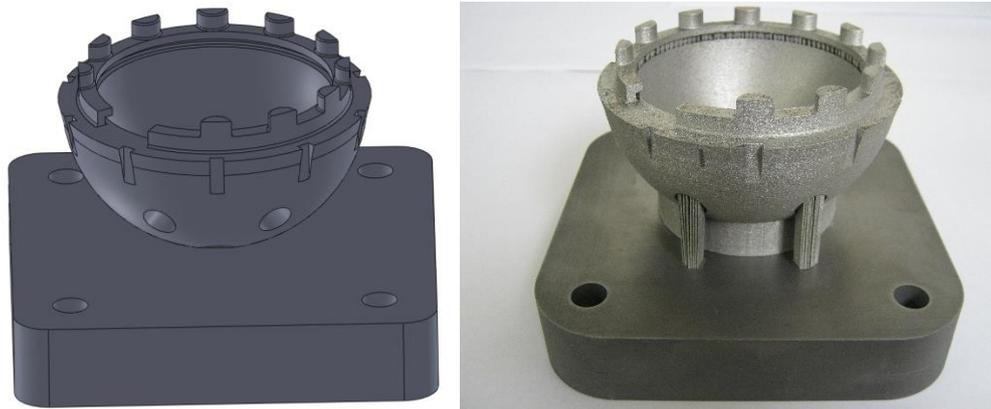

Figure 20 (Left) prosthetic acetabular hip designed by software (Right) produced sample by using SLM method (193)

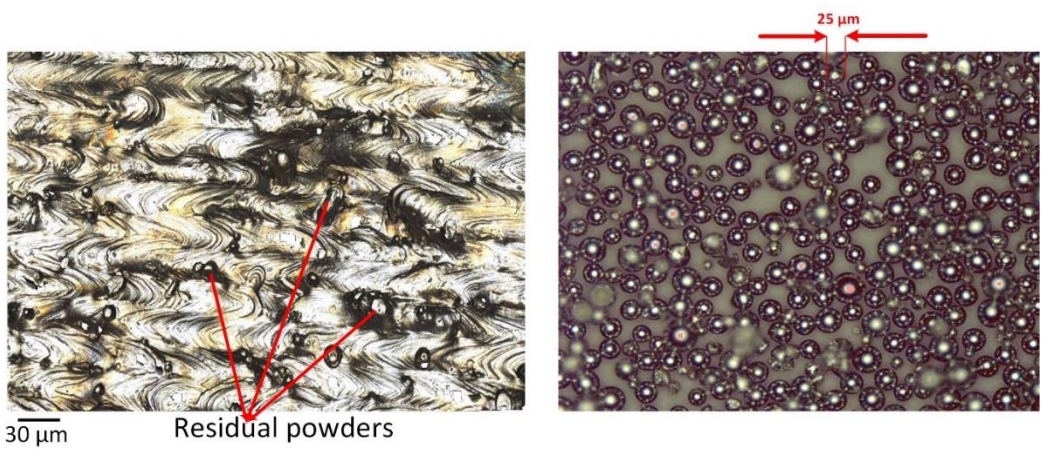

Figure 21 (Left) Powder particles on the surface of the sample in SLM methods (Right) Ti powder for SLM (193)

This method is very sensitive to initial conditions, expensive in run and powder particles remain on the surface after the production process. Figure 22 (left) and 23 show that the initial surface doesn't have great quality, however this method decreases manufacturing steps and time. Compatibility with rapid 3D design allows characterizing it as one of the promising methods in biofabrication. Figure 22 (right) shows the powder size that was used for producing prosthetic hip by using SLM method. Figure 23 shows good deviation between the produced sample from the SLM machine and base circle design, showing the high accuracy and versatility of this method for biofabrication.

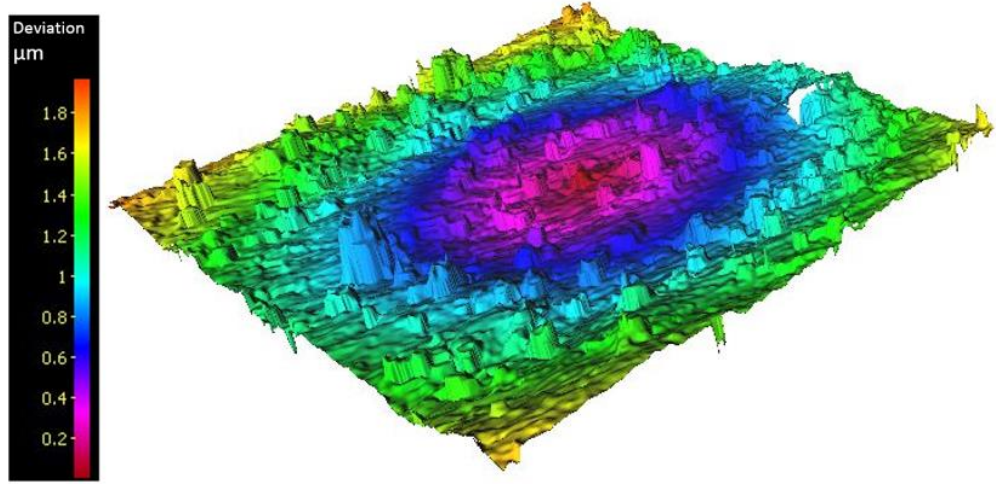

Figure 22 The inner surface of prosthetic acetabular produced by SLM (193)

### 1.10.4   Other fabrication methods

In this section other manufacturing methods for Ti-Based biomaterials such as superplastic forming, forging, ring rolling and joining methods will be discussed.

Superplastic property is expressed as having a strain rate sensitivity exponent of 0.5 or greater in the flow stress and strain rate. Sensitivity is computed as per Equation 2:

$$\sigma = \sigma_0 \, (d\varepsilon/dt)^m \hspace{3cm} (2)$$

where $\sigma_0$ is the threshold flow stress at very low strain rate and (m) is strain rate. In creep forming with strain rates of $10^{-5}$ to $10^{-6}$ s$^{-1}$ and a temperature of $0.6T_M$ ($T_M$ is melting temperature), if alloys have stable small grain size they have superplastic forming characteristics. In Ti-Based biomaterials Ti-6Al-4V and CP-Ti have superplastic forming capacity at a temperature of $950^0$C and pressure of 1Mpa. The ability to produce large and complex samples in one step, good surface finish, high accuracy and low deformation force are the main advantages of this method. In addition, no residual stress and spring back was observed in this operation, while oxidation of workpiece and equipment, slow forming rate and lower volume of production are the disadvantages of this approach in manufacturing of prosthetic biomaterials (34, 231-234).

Ti forging is performed in either open or close dies. The number of forging operations is related to the size of a production, complexity of the shape and workability of the alloy being forged. Preheating operations after passing each stage are recommended. In Ti forging basket-weave microstructure is observed in bars produced by β finish-forging method; while duplex or tri-modal microstructure with near β phase appears in bars manufactured by forging technology (235-237). In forging grains are continuous throughout the part so the strength is higher than other methods such as casting. In order to avoid undesirable work hardening hot forging is recommended. Work hardening in cold forging is not economical and causes difficulty on secondary processes like machining. Alloys with precipitation hardening capacity such as AL and Ti can be hot forged followed by a hardening step. In hot forging dangers are present due to high temperatures and pressures, in addition facilities and personnel are expensive. In cold forging higher flow stress and net shape memory characteristics lead to increase in the cost of the process. In hot forging the metal forming die and mold must be prepared accurately after machining and heat treatment which adds to the final production costs.

Ring rolling is used for the production of Ti cylindrical biomaterials specially CP-Ti and Ti-6Al-4V. In this process the crystal structure of productions is weakened by twinning, however strengthened by slip (238). Theoretically, the feasibility of cold rolling of Ti sheet under tensional forces mainly depends on the plastic deformation conditions and state of stress. Cold rolling needs expensive equipment due to the high hardness and tensile strength of Ti, however gives a good surface finish and dimensional deviations (239). A mix of twist extrusion and rolling of CP-Ti leads to an additional refinement of the microstructure, which remains thermally stable at 300–350°C (240).

Fusion welding and friction welding are the most common joining methods for Ti and Ti-Based biomaterials. "Typically, it becomes more difficult to produce welded structures, or products as the alloy strength increases. This is because the properties of the weld do not match those of the base metal and because some of the high strength alloys contain eutectoid alloying elements that impair the solidification, integrity, and properties of the welds. An additional issue is the availability of filler wire that matches the composition of the base metal". Friction welding of Ti is used to produce high integrity joints with α+β phase, the best method for welding of non-axisymmetric Ti and Ti-Based alloys is linear friction welding. Welding is not common in manufacturing of biomaterial and most of the prosthetic organs are produced by other methods (34, 241, 242).

### 1.10.5  HA-Ti bio-composite manufacturing

Fabrication HA-Ti bio-composites have good potential in biofabrication and their production has been carried out by different approaches such as sol-gel, thermal spraying, electrophoretic deposition, hot pressing and hot isostatic pressing, pulsed laser deposition, sputter coating and dip coating. Table 4 shows advantages and disadvantages of the mentioned methods for HA-Ti bio-composite fabrication.

Table 4 Techniques to produce HA-Ti bio-composite (243-251)

| Technique | Thickness | Advantages | Disadvantages |
|---|---|---|---|
| Dip coating | 0.05-0.5mm | Inexpensive to run | needs high sintering temperatures |
| | | Fast coating process | Thermal expansion misalliance |
| | | Impossible for intricate substrates | |
| Electrochemical deposition | 2µm | Union grains for as built samples | Crystal morphology and size are sensitive to electrolyte temperature |
| | | Thick layers by increasing deposition time | Crystal morphology and size are sensitive to process time |
| Electrophoretic deposition | 0.1-2.0mm | Uniform coating thickness | Cracks on the surfaces |
| | | High deposition rates | High sintering temperatures |
| | | Impossible for intricate substrates | |
| Hot Pressing and hot isostatic pressing | 0.2-2.0mm | High density in the surface | Impossible for intricate substrates |
| | | | High operation temperature |
| | | | Thermal expansion incompatibility |
| | | | Variation in elastic property |
| | | | Expensive to run |
| | | | Removal/Interaction of encapsulation material |
| Metal injection molding | NA | Reduce production costs | Contaminates |
| | | Homogeneity | High risk of cracks |
| | | Selectable mechanical properties by changing sintering temperature and cooling rate | Decomposition at temperature above $1100^0C$ |
| | | | Alternation in mechanical properties after implantation in SBF |
| Powder metallurgy | NA | Excellent microstructure | Expensive instruments |
| | | Near net shape productions | |

| Pulsed laser deposition | 0.05- 5µm | As for sputter coating | As for sputter coating |
|---|---|---|---|
| Sol-Gel | <1µm | Impossible for intricate substrates | Needs more equipment to control atmosphere processing |
| | | Low temperatures in operation | Expensive raw materials |
| | | Relatively inexpensive and thin | |
| Sputter coating | 0.02-1µm | Uniform coating thickness on flat substrates | Line of sight technique |
| | | | Expensive to run |
| | | | Slow operation |
| | | | Impossible for intricate substrates |
| | | | Amorphous productions |
| Thermal spraying | 30-200µm | High deposition rates | Line of sight technique |
| | | | High temperatures induce decomposition |
| | | | Thermal cooling gradient produces amorphous coatings |

## 1.11 Discussion and future work

Different problems in biomaterials in terms of material metallurgy, microstructure, fabrication methods, mechanical properties, corrosion, biocompatibility, surface modification and osseointegration lead to failure such as cracks, deformation and fracture. Therefore, revision surgery must be implemented to replace artificial implants that are time consuming and expensive. Generally speaking, biomaterial failure is a significant issue that is associated with, low wear and corrosion resistance, fibrous encapsulation, release of residual stress, low surface quality, decreasing in osseointegration, mismatching in bone and elastic modulus, low fatigue stress, low fracture toughness and inflammation. The most promising method in biofabrication is AM which can be used for production of different parts such as hip, heart valve, knee and dental implants. Indeed, the ability to choose porosity of the produced material in this technique is another exclusive property that can change modulus, hardness, corrosion resistance, cell adhesion, osseointegration and biocompatibility.

Future work will be dedicated to optimization of AM methods for achieving prosthetic parts with superior mechanical properties, biocompatibility, osseointegration and anti-corrosion characteristics in simulated human body fluid. Also, improving and enhancing the quality of Ti-Based biomaterials during manufacturing processes such as optimization in material removal processes or forming, heat treatment, oxidation, surface improvement, surface polishing and surface coating will be subjected to prospect attentions.

## 1.12 Conclusion

In this paper, Ti and Ti-Based alloys used in biomedical applications as well as different aspects such as metallurgy, mechanical properties, surface modification, anti-corrosion characteristics, biocompatibility and osseointegration have been discussed. Moreover, advantages and disadvantages for various Ti production processes in biomedical applications such as casting, powder metallurgy, cold and hot working, machining, additive manufacturing, superplastic forming, forging and ring rolling have been outlined. The most significant points for Ti and Ti-Based alloys in medical applications in terms of cytotoxity, mechanical and anti-corrosion characteristics and biocompatibility are presented below.

Ti-Based alloys in $\alpha+\beta$ phase have low toxicity and low allergenic properties, and the cytotoxicity ranking of various elements in Ti-Based biomaterials from the most to least is recognized as: Cu > Al=Ni > Ag > V > Mn > Cr > Zr > Nb > Mo >Ta>Sn> CP-Ti.

Heat treatment improves biomaterial characteristics for instance, aging increases tensile strength, brittleness, mechanical properties and biocompatibility. Annealing increases breaking elongation of Ti based biomaterials, however decreases yield and tensile strength.

Compared to copper and stainless steels Ti has higher corrosion resistance. Ageing, quenching, plastic deformation, formation of combination of Ti and Mo, porous layers on the surface, heat treatment and $\beta$ phase stabilizer elements improve Ti's characteristic on corrosion resistance. Some phases such as $\alpha+\beta$ and $\beta$ due to difficulty in initiating cracks have good anti-corrosion characteristics, but some $\beta$ phase stabilizers such as Mo are not suitable due to releasing to the surrounding tissue. Ti coating procedures such as plasma nitriding, CVD, PVD, ion nitriding production of oxynitrides, oxygen diffusion, diamond like coating, laser annealing and enrichment of nitrogen on passive layers increases the resistance of produced components against corrosion and wear. Moreover, dislocation density of Ti-Based alloys that is observed in machining or forming procedures changes surface hardness and micro hardness, which is called work hardening.

The ranking of cell viability of elements which are added to the Ti in bioimplants from the most to least strength enhancing is: CP-Ti>Mo> Nb>Zr>Cr>Mn>V>Ag>Al>Cu. Increasing oxide thickness and HA coating by using different methods such as plasma spraying, sol-gel, surface porosity, thermal heat treatment, surface grit blasting, polishing, antibacterial coating and biomimetic processes improve osseointegration and biocompatibility of Ti-Based biomaterials. Furthermore, better characteristics such as uniform density, structure homogeneity, well crystallized structure and anti-corrosion properties are obtained by using HA coating in Ti-Based biomaterials. Grooved surfaces help promote osteoblastic cell attachment, proliferation and adhesion therefore, osseointegration and biocompatibility are improved.

Because of high strength, low thermal conductivity, shape memory, high hardness and spring back, Ti-Based alloys in biomaterial has low capacity in cold working and machining. Choosing the porosity value of productions in AM changes cell adhesion, osseointegration, osteoconduction, biocompatibility and hardness that allows characterizing AM as a new and efficient method with high flexibility in fabrication of complicated prosthetic organs such as heart valve, hip and knee.